\documentclass[11pt,final,reqno]{amsart}
\newcommand{\normalspacing}{\renewcommand{\baselinestretch}{1.1}\tiny\normalsize}

\normalspacing

\usepackage{amsmath,amsfonts,amssymb,alltt,verbatim,xspace}
\newcommand{\Vfile}[1]{ \scriptsize\begin{quote}\rule{4.6in}{0.1mm} \verbatiminput{#1} \rule{4.6in}{0.1mm}\end{quote}\normalsize }
\newcommand{\Vend}{ \rule{4.6in}{0.1mm}\normalsize\end{quote} }
\newcommand{\beginV}{ \begin{quote}\rule{4.6in}{0.1mm}\scriptsize\begin{verbatim} }

\usepackage{harvard} 
\usepackage[final]{graphicx}
\newcommand{\widefigure}[2]{\includegraphics[keepaspectratio=true,
width=#2in]{figs/#1.eps}}

\newtheorem*{lem}{Lemma}
\theoremstyle{definition}
\newtheorem*{defn}{Definition}

\newcommand{\br}{\mathbf{r}}

\newcommand{\Div}{\nabla\cdot}

\newcommand{\ppr}[1]{\frac{\partial #1}{\partial r}}
\newcommand{\ppt}[1]{\frac{\partial #1}{\partial t}}

\newcommand{\ppxx}[1]{\frac{\partial^2 #1}{\partial x^2}}
\newcommand{\ppyy}[1]{\frac{\partial^2 #1}{\partial y^2}}

\newcommand{\grad}{\nabla}

\newcommand{\kap}{\kappa}
\newcommand{\lam}{\lambda}
\newcommand{\lap}{\triangle}
\newcommand{\mtt}{\texttt}
\newcommand{\Matlab}{\textsc{Matlab}\xspace}
\newcommand{\RR}{\mathbb{R}}

\newcommand{\vf}{\varphi}

\newcommand{\Bueleretal}{(Bueler et al.~2005)\nocite{BLKCB}}

\title[Computation of a viscoelastic deformable earth model]{Computation of a combined spherical-elastic and viscous-half-space earth model for ice sheet simulation}

\author[Bueler and others]{Ed Bueler$^1$, Craig S. Lingle$^2$, and Jed A. Kallen-Brown$^1$}

\thanks{\!\!\!\!\!\!\!\textsc{Draft \today.} \\  $^1$Dept.~of Mathematics and Statistics, Univ.~of Alaska, Fairbanks AK 99775-6660.  Email \texttt{ffelb\@@uaf.edu}. \\ $^2$Geophysical Institute, Univ.~of Alaska, Fairbanks AK 99775-6660.}

\begin{document}

\begin{abstract}  This report starts by describing the continuum model used by \citeasnoun{LingleClark} to approximate the deformation of the earth under changing ice sheet and ocean loads.  That source considers a single ice stream, but we apply their underlying model to continent-scale ice sheet simulation.  Their model combines \possessivecite{Farrell} elastic spherical earth with a viscous half-space overlain by an elastic plate lithosphere.  The latter half-space model is derivable from calculations by \citeasnoun{Cathles}.  For the elastic spherical earth we use Farrell's tabulated Green's function, as do Lingle \& Clark.  For the half-space model, however, we propose and implement a significantly faster numerical strategy, a spectral collocation method \cite{Trefethen} based directly on the Fast Fourier Transform.  To verify this method we compare to an integral formula for a disc load.  To compare earth models we build an accumulation history from a growing similarity solution from (Bueler, et al.~2005)\nocite{BLKCB} and and simulate the coupled (ice flow)-(earth deformation) system.  In the case of simple isostasy the exact solution to this system is known.  We demonstrate that the magnitudes of numerical errors made in approximating the ice-earth system are significantly smaller than pairwise differences between several earth models, namely, simple isostasy, the current standard model used in ice sheet simulation \cite{Greve2001,HagdornThesis,ZweckHuybrechts}, and the Lingle \& Clark model.  Therefore further efforts to validate different earth models used in ice sheet simulations are, not surprisingly, worthwhile.\end{abstract}

\maketitle
\thispagestyle{empty}


\section{Two linear earth models and their Green's functions}\label{sect:twolin}

\citeasnoun{LingleClark} use as their fundamental tools the Green's functions of two different linear earth models.  The Green's functions for these models are convolved with the load to compute (vertical) displacements of the earth's surface.  One finds an elastic displacement $u^E$ and a viscous displacement $u^V$ given a current load and a load history, respectively, as we will explain.  The total displacement is then the sum $u=u^E+u^V$ at any time.  That is, the two linear models are superposed.

The partial differential equations (PDEs) behind these Green's functions are linear.  In this report we state these PDEs, which is, in the case of the second model, a nontrivial accomplishment (see section \ref{sect:PDEmethod}).  We then approximately solve these PDEs in a demonstrably efficient manner.  First, however, we describe the two models and their sources in the literature.

\newcommand{\bs}{\,\mathbf{s}}
\newcommand{\bu}{\mathbf{u}}
\newcommand{\er}{\mathbf{e}_r}
\newcommand{\etheta}{\mathbf{e}_\theta}
\subsection*{An elastic, self-gravitating spherical earth}  The main equations of this model are labeled (27) in \cite{Farrell}:
\begin{gather*}
\Div\tau - \grad\left(\rho g \bs\cdot\er\right) -\rho\grad\phi + g \Div\left(\rho\bs\right)\er + \rho \ddot \bs = 0, \\
\grad^2 \phi = - 4\pi G \Div\left(\rho\bs\right)
\end{gather*}
Here $\rho=\rho(r)$ is the density of the earth.  (We restrict to only radial dependence because the densities used by \cite{Farrell} are for ``stratified'' earths.)  Also, $g$ is the acceleration of gravity, $\tau$ is the full stress tensor, and $\bs$ is the displacement vector field (i.e.~the strain field) which we seek.  The gravitational potential $\phi$ is described below.  Note that $\dot \bs = \bu$ is the velocity field and that $\ddot \bs$ is just the acceleration.

The first equation comes from formally linearizing the equation of conservation of momentum.  The second equation is for the gravitational potential.  The field $\phi$ is the \emph{additional} gravitational potential ``on top of'' that caused by the undeformed earth and also additional to the potential of any masses outside the earth (including the load).  In Farrell's words ``\dots $\phi$ is the perturbation in the ambient graviational potential $\phi_1$ plus the potential of any externally applied graviational force field $\phi_2$.''

Actually, equation (27) in \cite{Farrell} has ``$-\omega^2 \rho\bs$'' where we have ``$\rho \ddot \bs$'' because Farrell only states the Fourier-transformed-in-time equations.  We will only be interested in the $\ddot \bs=0$, equivalently $\omega=0$ case, however, because we are interested in phenomena on the scale of years or centuries, unlike Farrell who was interested in tides.

Consider point forces or disk loads at the surface.  For such loads it is natural to use spherical coordinates $(r,\theta,\vf)$ where the $z$-axis is along the line between the center of the earth and the center of the load.  Here $\theta$ is the angle between the position vector and the $z$-axis  (i.e.~the colatitude) and $\vf$ is the longitude.  The load is at the ``north pole.''

For the resulting stratified problem we seek the components of $\bs$ which do not vanish, namely $s_r$ and $s_\theta$ in $\bs = s_r(r,\theta)\er + s_\theta(r,\theta)\etheta$.  By symmetry the ``toroidal'' component of the strain, $s_\vf$, vanishes everywhere.  Also we seek the potential $\phi=\phi(r,\theta)$.  The functions $s_r,s_\theta,\phi$ are expanded in spherical harmonics with radially-dependent coefficients; see equation (28) in \cite{Farrell}.  The radially-dependent coefficients, for each degree in the expansion, solve a system of ODEs in the radial coordinate $r$.  Using a radially-dependent density for the earth these can be solved numerically by standard ODE means.  Note that \cite{Farrell} and \cite{LingleClark} choose the ``Gutenberg-Bullen A'' model.  For the Green's function corresponding to a point load, Farrell has done this using a Runge-Kutta method, and we accept and use the tabulated result.

Following \cite{LingleClark} we are only interested in the vertical displacement of the surface of the earth, and therefore the vertical displacement $u(\theta)$ corresponding to a point load is the Green's function we seek.  In terms of the spherical harmonics expansion the relevant equation is equation (37) in \cite{Farrell}.  Farrell computes this Green's function and reports its values at particular distances in his table A3.  Table 1 in \cite{LingleClark} also reports this data.  One must be clear on normalization so a plot is in order here.  Let $G^E(r)$ be the vertical displacement caused by a 1 kg mass at the north pole and evaluated at a distance $r$ along the surface of the earth.  (The coordinate $r$ here has a different meaning from the spherical coordinate of the same name.  Specifically, the new variable $r=a\theta$ if $a$ is the radius of the earth and $\theta$ is the spherical coordinate, the radian colatitude.)  Figure \ref{fig:GEgraph} shows $G^E(r)$.  There is a $1/r$ singularity to this elastic Green's function, in contrast to the Green's function for the flat, viscous model which follows.

\begin{figure}[ht]
\centerline{\widefigure{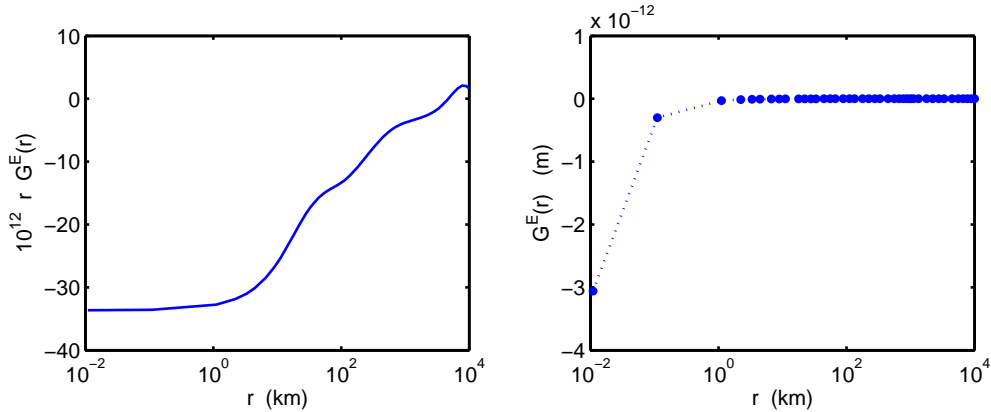}{6}}
\caption{Two views of the vertical surface displacement Green's function $G^E(r)$ for the elastic spherical self-gravitating earth model (Farrell 1972).  Here $r$ is the distance along the surface of the earth from the point of application of the load.  Left: the smooth normalized form $r G^E(r)$.  Right: the same data without normalization, suggesting the actual $1/r$ singularity.  Note log scale on the horizontal axes.} \label{fig:GEgraph}
\end{figure}

This elastic Green's function is used as described in equation (20) in \cite{LingleClark} and as follows.  Suppose we seek the vertical displacement $u^E=u^E(x,y,t)$, caused by elastic deformation of the spherical earth.  Suppose the load at time $t$ is given by the function $\Psi(x,y,t)$, with units of mass per unit area.  Then
\begin{equation}\label{GEuse}
u^E(x,y,t) = \iint\limits_R G^E(|\br-\br'|)\Psi(x',y',t)\,dx' dy',
\end{equation}
where we define $|\br-\br'|^2=(x-x')^2+(y-y')^2$, of course, and where $R$ denotes a map-plane region containing the load.  Clearly the displacement $u^E$ depends on time only through the changing load; elastic changes are instantaneous.

In using \eqref{GEuse} we necessarily project the earth's geoid into a fixed plane.  This projection means our results are limited to an appropriately small region of the earth's surface.  We do integral \eqref{GEuse} numerically as explained in section \ref{sect:computelr}.

\subsection*{A viscous, flat earth overlain by an elastic plate}  Next we describe the time-dependent Green's function for a model which comes from \cite{Cathles}.  The PDE actually solved by this Green's function is given in section \ref{sect:PDEmethod}.

Cathles' sub-subsection III.A.2.e, pp.~50--55, describes a viscous half-space asthenosphere overlain by an elastic plate lithosphere.  An important point about this model, which partly explains the superposition ``$u^E+u^V$'' used by \cite{LingleClark}, is that the elastic plate lithosphere used here deflects but does not compress in the vertical.  Therefore all vertical motion in this model is really asthenosphere motion, though the elastic plate spreads the influence of any load.  The just-described spherical elastic earth exhibits elastic compression, however.

Unfortunately, all we are given in \cite{Cathles} are the Hankel transforms of the actual equations.  Roughly speaking, from \cite{Cathles} we use equation (III-35) along with the definitions of ``$\alpha$'' and ``$D$'' contained in the footnote on page 52 (Lingle 2005, personal communication).  Precisely speaking, however, our source is \cite{LingleClark}, from which we use equations (4), (7), and (8)--(14).  We will also use the particular choices of layer thickness, viscosity, and flexural rigidity for the ``two-layer'' model from that source.  We now repeat some equations from \cite{LingleClark} as needed for clarity.

\newcommand{\baruV}{{\bar u}^V}
\newcommand{\barUV}{{\bar U}^V}
Let $u^V(r,t)$ be the vertical displacement of the surface supposing a point load at the origin $r=0$ applied at time $t=0$ and held.  Consider the Hankel transform of this function
    $$\baruV(\kappa,t)= \int_0^\infty u^V(r,t) J_0(\kappa r) r\,dr,$$
where $J_0$ is the Bessel function of zero order (see Appendix \ref{app:hankel}).  The Hankel transform is self-inverse, so $u^V$ can be recovered from $\baruV$ by the same integral.

The half-space model hypothesizes \cite{LingleClark} that $\baruV$ solves the equation
\begin{equation}\label{cathlespde}
\ppt{\baruV} + \frac{\rho_r g\alpha(\kappa)}{2\eta\kappa} \baruV = \frac{{\bar \sigma}_{zz}(\kappa,t)}{2\eta \kappa}
\end{equation}
where
    $$\alpha(\kappa)=1+\frac{D \kappa^4}{\rho_r g} \quad \text{ and} \quad D=\frac{E T^3}{12(1-\nu^2)}.$$
We denote by ${\bar \sigma}_{zz}$ the Hankel transform of the normal stress from a point load applied at the origin
\begin{equation}\label{cathlesload}
{\bar \sigma}_{zz}(\kappa,t) = -\frac{g}{2\pi} H(t),
\end{equation}
in units of $\text{N}\, \text{m}^{-2}$, corresponding to a point mass of $1$ kg.  Here $H(t)$ is the Heaviside function ($H(t)=1$ for $t\ge 0$ and $H(t)=0$ otherwise).  Note ${\bar \sigma}_{zz}(\kappa,t)$ is the Hankel transform of $\sigma_{zz}(r,t)=-g \delta_0(r) H(t)$ where $\delta_0(r)$ is the Dirac delta function at the origin which acts on functions on the plane.\footnote{See Appendix \ref{app:hankel} and especially equation \eqref{delta0defn} for the defining property of $\delta_0(r)$.}  The initial condition to \eqref{cathlespde} is the condition of zero displacement
    $$\baruV(\kappa,0)=0.$$
That is, $\baruV(\kap,t)$ is the Hankel transform of the Heaviside Green's function of an as yet unstated PDE of which \eqref{cathlespde} is the Hankel-transformed version; see section 3 for clarification of this description.

Poisson's ratio and Young's modulus for the elastic plate lithosphere are assumed to be $\nu=0.5$ and $E=6.6\times 10^{10} \, \text{M}/\text{m}^2$, respectively.  The lithosphere thickness $T$ is assumed to be $88\,\text{km}$.  The resulting flexural rigidity is $D=5.0\times 10^{24}$ N m.  The density and viscosity of the fluid in the underlying half-space are assumed to be $\rho_r=3300\,\text{kg}/\text{m}^3$ and $\eta=10^{21}\,\text{Pa}\,\text{s}$, respectively.

Equation \eqref{cathlespde} is an uncoupled set of linear first order ODEs in time.  That is, the spatial Hankel transform has done its job and turned a PDE into a solvable system.  Let\footnote{Use of $\beta$ instead of $\alpha$ represents an admittedly minor simplification of the notation in \cite{LingleClark}.} $\beta(\kap)=\rho_r g \alpha(\kap)=\rho_r g + D \kap^4$.  The solution of \eqref{cathlespde} and \eqref{cathlesload} is
\begin{equation}\label{barghv}
\baruV(\kappa,t)=-\frac{g}{2\pi\beta(\kappa)} \Big(1-\exp\left[-\beta(\kappa) t/(2\eta \kappa)\right]\Big)
\end{equation}
for $t>0$ and $\baruV(\kap,t)=0$ for $t\le 0$.  Because of the self-inverse property of the Hankel transform, we have the following integral formula for the Green's function: \begin{equation}\label{ghv}
G^V(r,t)=u^V(r,t)=-\frac{g}{2\pi} \int_0^\infty \beta(\kappa)^{-1} \Big(1-\exp\left[-\beta(\kappa) t/(2\eta \kappa)\right]\Big)\,J_0(r\kappa)\,\kappa\,d\kappa,
\end{equation}
for $t>0$ and $G^V(r,t)=0$ for $t<0$.  This formula is equation (14) in \cite{LingleClark}.  Note $G^V$ has units $\text{m kg}^{-1}$; see formula \eqref{GHVuse} below.

As far as we know the integral \eqref{ghv} must be computed numerically.  Furthermore there seems to be no one-dimensional procedure analogous to the Fast Fourier Transform (FFT) \cite{Bracewell} to do the job quickly.  Our strategy for the similar disc load integral (Appendix \ref{app:exactdisc}) is to break up the oscillatory integral into more than 100 subintervals and call an adaptive quadrature routine for each subinterval.  This strategy is essentially the same as that described on page 1104 of \cite{LingleClark} for \eqref{ghv}.

Unlike the elastic case, the Green's function $G^V$ is time-dependent.  A graph of $G^V$ for several $t$ values is shown in figure \ref{fig:ghv}.   The viscous behavior is clear, as is the role of the elastic plate lithosphere in removing any singularity at $r=0$.  Note that the peripheral bulge develops only at large times; compare the ``standard'' model in section \ref{sect:discuss}.

\begin{figure}[ht]
\widefigure{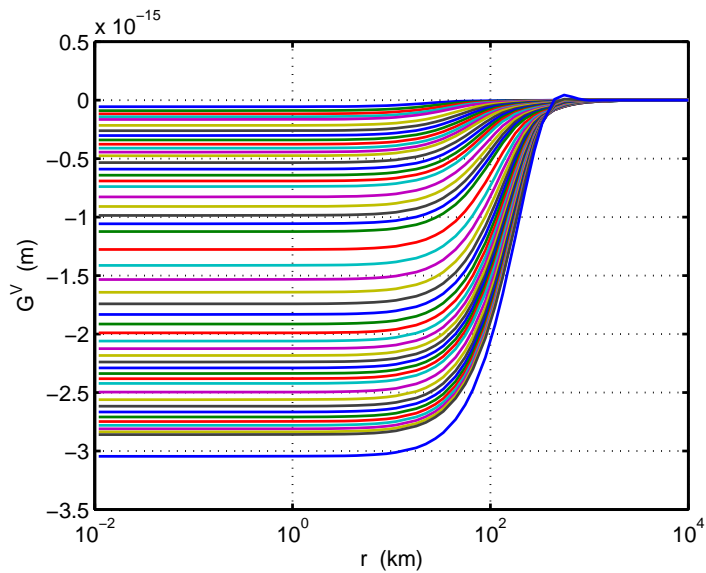}{4}
\caption{Green's function $G^{V}(r,t)$ for the viscous flat earth model.  The curves are from times given in table 1 in (Lingle \& Clark 1985); compare figure figure 5(a) there.  The top curve is at $t=20$ years and the bottom at $10^5$ years.  Not log scale on the horizontal axis.} \label{fig:ghv}
\end{figure}

A method for using $G^V$ to compute the response to arbitrary load is described in equations (19) and (22) in \cite{LingleClark}, and as follows.  Suppose there is a load function $\Psi(x,y,t)$, with units of mass per unit area, on some region $R$ of our map-plane.  The time rate of change of this load function, or, equivalently, the incremental changes in this load function, are what we sum using $G^V$ to find the general (viscous) displacement $u^V=u^V(x,y,t)$ of the surface.  In fact, let
    $$\lam(x,y,t) = \ppt{\Psi}(x,y,t).$$
Then Lingle \& Clark assert (with a heuristic motivation; pp.~1105--1106) that:
\begin{equation}\label{GHVuse}
u^V(x,y,t) = \int_{-\infty}^t \iint\limits_R G^V(|\br-\br'|,t-t') \lam(x',y',t') \,dx'\,dy'\,dt'.
\end{equation}
Note that the right side of \eqref{GHVuse} has units of length because $G^V$ has units $\text{m}\,\text{kg}^{-1}$, as noted, and $\lam$ has units $\text{kg}\,\text{m}^{-2}\,\text{s}^{-1}$.

\citeasnoun{LingleClark} describe the function $\lam$ discretely, thereby incorporating an approximation of the rate of change of the load.  Some such approximation is essential in practice, of course.  They suppose a fixed sequence of past times $\{t_i\}$ and define
    $$\Lambda(x,y,t_i) = \int_{t_{i-1}}^{t_i} \lam(x,y,t)\,dt = \Psi(x,y,t_i)-\Psi(x,y,t_{i-1}),$$
so
\begin{equation}\label{LCuse}
u^V(x,y,t) \approx \sum_i \iint\limits_R G^V(|\br-\br'|,t-t_i) \Lambda(x',y',t_i) \,dx'\,dy';
\end{equation}
compare equations (16) and (17) in \cite{LingleClark}.

\section{The load response matrix method}\label{sect:computelr}

Suppose now that the map-plane region $R$ is divided into a grid of rectangular elements of area $\Delta x\Delta y$.  Concretely, suppose $R=[-L_x,L_x]\times[-L_y,L_y]$ is a rectangular region, suppose $N_x,N_y$ are positive integers, and let $\Delta x = (2 L_x)/N_x$, $\Delta y=(2L_y)/N_y$.  Center each element at $(x_j,y_k)=(-L_x+(j-1/2)\Delta x,-L_y+(k-1/2)\Delta y)$.  There are $M=N_x N_y$ elements, each denoted by a pair $(j,k)$ for $1\le j \le N_x$, $1\le k \le N_y$.

We define the \emph{elastic load response matrix} (LRM) $\left\{\Gamma_{(jk)(mn)}\right\}$, $j,m=1,\dots,N_x$, $k,n=1,\dots,N_y$, as the vertical displacement of element $(j,k)$ caused by a unit change in ice thickness within element $(m,n)$; compare \cite{LingleClark} page 1106.  This displacement is assumed constant within element $(j,k)$.  (As it stands, this description applies only to the elastic spherical model in the previous section.  Slightly different conventions apply to the LRM for the viscous earth model; see below.)

We compute $\left\{\Gamma_{(jk)(mn)}\right\}$ by integrating over element $(m,n)$:
\begin{equation}\label{Gammaintegral}
\Gamma_{(jk)(mn)} = \rho_i \int_{y_n-\Delta y/2}^{y_n+\Delta y/2} \int_{x_m-\Delta x/2}^{x_m+\Delta x/2} G^E\left(\sqrt{(x_j-x)^2+(y_k-y)^2}\right)\,dx\,dy,
\end{equation}
where $\rho_i=910 \text{ kg } \text{m}^{-3}$ is the density of ice.  Note that the load per unit area (i.e.~$\Psi$ in the last section) for a unit (i.e.~one meter) change in ice thickness on a rectangle of area $\Delta x\Delta y$ is
    $$\Psi=\frac{\text{mass}}{\text{area}} = \frac{\rho_i(1\cdot \Delta x\cdot \Delta y)}{\Delta x\Delta y} = \rho_i.$$
That is, equation \eqref{Gammaintegral} is a special case of equation \eqref{GEuse}.

The matrix $\left\{\Gamma_{(jk)(mn)}\right\}$ is $M\times M$ if we linearly order all elements; alternatively $\Gamma$ could be regarded as a ``$4$-tensor'' with indices $j,k,m,n$.   There are symmetries in this object, and we may exploit them to compute roughly $(2N_x)(2N_y)=4M$ integrals \eqref{Gammaintegral} rather than doing an integral for each of the $M^2$ entries of $\Gamma_{(jk)(mn)}$.  In fact, if we change variables in \eqref{Gammaintegral} by $x=x_m-\xi$, $y=y_n-\zeta$ then we get
    $$\Gamma_{(jk)(mn)} = \rho_i \int_{-\Delta y/2}^{\Delta y/2} \int_{-\Delta x/2}^{\Delta x/2} G^E\left(\sqrt{((j-m)\Delta x - \xi)^2+ ((k-n)\Delta y-\zeta)^2} \right)\,d\xi\,d\zeta.$$
Let
\begin{equation}\label{Ipq}
I^E(p,q)=\int_{-\Delta y/2}^{\Delta y/2} \int_{-\Delta x/2}^{\Delta x/2} G^E\left(\sqrt{(p\Delta x - \xi)^2+ (q\Delta y-\zeta)^2} \right)\,d\xi\,d\zeta
\end{equation}
for $-N_x+1\le p \le N_x-1$, $-N_y+1\le q\le N_y-1$.  Then
    $$\Gamma_{(jk)(mn)} = \rho_i I^E(j-m,k-n).$$
We need only compute the $(2N_x-1)(2N_y-1)$ entries of $I^E$.  Integral \eqref{Ipq} can be done numerically, getting values for the integrand $G^E(\cdots)$ by interpolation between the values computed by Farrell.

Now let $H_{(mn)}$ be the average value of the ice thickness $H(x,y,t)$ over element $(m,n)$.  Following Lingle \& Clark we call $\left\{H_{(mn)}\right\}_{m,n=1}^{N_x,N_y}$ the \emph{load vector}. (Technically it is a \emph{thickness} vector, actually; if the load is actually liquid water one computes the equivalent thickness to give values $H_{(mn)}$.)  Integral \eqref{GEuse}, which gives the elastic displacement from the load, is approximated by
\begin{equation}\label{uEapprox}
u^E(x_j,y_k,t) \approx \sum_{m=1}^{N_x} \sum_{n=1}^{N_y} \Gamma_{(jk)(mn)} H_{(mn)} = \sum_{m=1}^{N_x} \sum_{n=1}^{N_y} \rho_i\, I^E(j-m,k-n)\, H_{(mn)}.
\end{equation}
Note that as a straightforward matrix-vector product, \eqref{uEapprox} requires $M=N_x N_y$ scalar multiplications to compute the elastic displacement in the $(j,k)$ element, and thus $M^2$ multiplications are required to update $u^E$ at each timestep.

A comparable LRM approximation applies to the viscous model.  Let $\{\Delta t_i\}_{i=1}^{N_t}$ be a decreasing sequence of $N_t$ positive times.  These values are time intervals before the present time; table 3 in \cite{LingleClark} gives a list of 26 values.  Let $\Phi_{(jk)(mn)}^i$ be the displacement in element $(j,k)$ caused by a change in ice thickness of $1$ meter in element $(m,n)$ at time $\Delta t_i$ before the current time.  From \eqref{LCuse},
\begin{equation}\label{PhiLRM}
\Phi_{(jk)(mn)}^i = \rho_i \int_{y_n-\Delta y/2}^{y_n+\Delta y/2} \int_{x_m-\Delta x/2}^{x_m+\Delta x/2} G^V\left(\sqrt{(x_j-x)^2+(y_k-y)^2},\Delta t_i\right)\,dx\,dy.
\end{equation}
Compare to equation \eqref{Gammaintegral} for the elastic case.  Let $\Lambda_{(mn)}^i$ be the average value of the change in ice thickness $H(x,y,t-\Delta t_{i+1})-H(x,y,t-\Delta t_i)$ over element $(m,n)$.  Then
\begin{equation}\label{uVapprox}
u^V(x_j,y_k,t) \approx \sum_{i=1}^{N_t} \sum_{m=1}^{N_x} \sum_{n=1}^{N_y} \Phi_{(jk)(mn)}^i \Lambda_{(mn)}^i.
\end{equation}
As it stands, matrix-vector product \eqref{uVapprox} requires $N_t M = N_t N_x N_y$ scalar multiplications to compute the $(j,k)$ element, thus $N_t M^2$ multiplications to update $u^V$.

As noted at the beginning, we superpose the results from these elastic and viscous LRM approximations:
\begin{equation}\label{LRMresult}
u(x_j,y_k,t) \approx \sum_{m,n} \Gamma_{(jk)(mn)} H_{(mn)} + \sum_i \sum_{m,n} \Phi_{(jk)(mn)}^i \Lambda_{(mn)}^i.
\end{equation}
Compare equation (25) in \cite{LingleClark}.

We are concerned with computability in reasonable time.  By using \eqref{LRMresult} to update every element at a given simulation timestep requires
\begin{equation}\label{LRMest}
(1+N_t) \cdot M \cdot M = O(N_t M^2) = O(N_t N_x^2 N_y^2)
\end{equation}
scalar multiplications.  \citeasnoun{LingleClark} used $M=38$, as this was the number of $50$ km long elements into which the flowline---a single ice stream and included ocean---was divided.  For Antarctica simulations with $\Delta x$ on the order of $50$ km or so, the minimum reasonable number of elements is $N_x=N_y$ at least $80$ so $M\ge 6400$.  Note that $N_t$ is roughly independent of the nature of the problem, as long as it involves large amounts of polar ice.  Thus by \eqref{LRMest} an Antarctica problem is roughly a factor of
    $$(6400^2)/(38^2) \approx 3\times 10^4$$
times more expensive in computation than the problem addressed in \cite{LingleClark} if one directly implements the load response matrix method using matrix multiplication.  Such poor scaling of this numerical method clearly represents a danger when using it in an ice sheet simulation.

A major speedup is possible if one uses the convolution sum form of the multiplication, as in equation \eqref{uEapprox} when $I^E$ is used.  (An obvious corresponding construction of a function $I^V(p,q,t)$ is needed to make the viscous LRM computations above into convolutions.  As it turns out, we will not need that construction.)  Convolutions sums can be quickly computed by the Fast Fourier Transform (FFT).  Using standard estimates on the time for the FFT \cite{BriggsHenson}, we can reduce the time to
\begin{equation}\label{LRMestwFFT}
(1+N_t) \cdot M \log M = O(N_t N_x N_y \log N_x \log N_y).
\end{equation}
Compare this with \eqref{LRMest} above.  One must still, however, precompute the LRMs, which turns out to be more expensive than solving the whole problem (over quite long time scales) if one uses the method of the next section.  In addition, the method of the next section eliminates a factor of $N_t$ work in a time-dependent simulation.  The method of the next section also significantly reduces memory usage.

\section{The straight-from-the-PDE method}\label{sect:PDEmethod}

\subsection*{Derivation}  We now reverse engineer some of the Green's function and Hankel transform ``thinking'' in the previous sections.  We recover the PDE underlying the half-space viscous model \eqref{cathlespde}.  Actually, the resulting equation is not, technically, a \emph{partial differential} equation.  It is a linear pseudo-differential equation easily understood through the Fourier transform.  We have already computed solutions of this PDE by the Hankel transform, by the indirect method of Green's functions.  In any case, analyzing the new PDE will lead to a much more efficient method for computing deformation in the half-space model.  We must still use, for now, the Green's function and LRM for the \emph{elastic} response computed from the \citeasnoun{Farrell} spherical earth model; we will implement the convolution sum \eqref{uEapprox} by the FFT.

Returning to equation \eqref{cathlespde}, we apply the inverse Hankel transform.  In fact \eqref{cathlespde} is equivalent to
\begin{equation}\label{hankpde}
\ppt{}\left(2\eta\kap\bar u\right) + \rho_r g \bar u +D\kap^4 \bar u = {\bar \sigma}_{zz},
\end{equation}
denoting $u=u^V$ for the rest of this section, and with the Hankel transform $\bar u=\baruV$.  As shown in Appendix \ref{app:hankel}, the multiplication by $\kap^1$ and $\kap^4$ which appear in equation \eqref{hankpde} can be regarded as the action of operators which are powers of the Laplacian operator.  In particular,
    $$\lap=-\grad^2=-\left(\ppxx{}+\ppyy{}\right)$$
acts on the Hankel transform of a radial function $f=f(r)$ by multiplication by $\kap^2$:
    $$\overline{\lap f}(\kap) = \kap^2 \, \bar f(\kap).$$
In fact $\lap$ is the \emph{positive} Laplacian as we see it is equivalent to multiplication by a nonnegative factor.  The inverse Hankel transform of \eqref{hankpde} is
\begin{equation}\label{pde}
\ppt{}\left(2\eta\,\lap^{1/2}\,u\right) + \rho_r g u +D\lap^2 u = \sigma_{zz}
\end{equation}
for $u(r,t)$.  Equation \eqref{pde} is the ``underlying PDE'' for equation \eqref{cathlespde}.  The symbol $\lap^2$ stands for the standard biharmonic fourth-order differential operator
    $$\lap^2 f = f_{xxxx} + 2 f_{xxyy} + f_{yyyy}$$
\cite[section 20]{Sneddon}.  The operator $\lap^{1/2}$ is not a differential operator but is definable via the Fourier transform in general; see Appendix \ref{app:hankel}.  One can also write equation \eqref{pde} as
    $$\ppt{}\left(2\eta\,|\grad|\,u\right) + \rho_r g u +D\grad^4 u = \sigma_{zz}$$
if the meaning $|\grad|=\sqrt{-\grad^2}$ is understood.

To confirm the equivalence of \eqref{cathlespde} and \eqref{pde} the reader may verify that the Green's function $G^V(r,t)$ defined by \eqref{ghv} satisfies
    $$\ppt{}\left(2\eta \lap^{1/2} G^V\right) + \rho_r g G^V + D \lap^2 G^V = - g \delta_0(r) H(t).$$

From now on we remove the assumption of radial load, and suppose equation \eqref{pde} applies for any load $\sigma_{zz}(x,y,t)$.  The solution $u(x,y,t)$ is a function of three variables; it is no longer radial.

Note that the equilibrium of \eqref{pde} is a standard rigid plate equation with a bouyant restoring force:
\begin{equation}\label{equilbouyant}
D\grad^4 u = \sigma_{zz}-\rho_r g u.
\end{equation}
For example, this is equation (8.7.3) in \cite{vanderVeen}.

The interesting part of \eqref{pde} is the time-derivative term.  This term accounts for viscous flow within the mantle.  It is not completely clear to the author at the present time why the particular power $\lap^{1/2}$ appears or why it represents the correct diffusive behavior.  The units in equation \eqref{pde} are consistent only with the $1/2$ power of the Laplacian, however.

The Fourier transform of \eqref{pde} is worth noting, because, as in the special case of a radial load wherein we may use the Hankel transform, we can write the solution as an integral.  Namely, if $\tilde u(\xi,\zeta,t)$ is the two-variable, spatial Fourier transform of $u(x,y,t)$,
    $$\tilde u(\xi,\zeta,t) = \frac{1}{2\pi} \int_{-\infty}^\infty \int_{-\infty}^\infty u(x,y,t) e^{-i (x\xi + y\zeta)}\,dx\,dy,$$
then \eqref{pde} is equivalent to
\begin{equation}\label{fourierpde}
\ppt{}\left(2\eta\,\left(\xi^2+\zeta^2\right)^{1/2}\,\tilde u\right) + \rho_r g \tilde u +D\left(\xi^2+\zeta^2\right)^2 \tilde u = {\tilde \sigma}_{zz}.
\end{equation}
As with equation \eqref{cathlespde}, this is a decoupled system of first order ODEs in time.  Using initial condition $u(x,y,t_0)$ the solution is
\begin{align}
\tilde u(\xi,\zeta,t) &= \int_{t_0}^t \frac{\exp\left[-\beta(\xi,\zeta)(t-s) /(2\eta(\xi^2+\zeta^2)^{1/2})\right]}{2\eta (\xi^2+\zeta^2)^{1/2}} \, \tilde \sigma_{zz}(\xi,\zeta,s)\,ds \label{fouriersoln} \\
    &\qquad\qquad + \exp\left[-\beta(\xi,\zeta)(t-t_0)/(2\eta(\xi^2+\zeta^2)^{1/2})\right] \tilde u(\xi,\zeta,t_0)\notag
\end{align}
where $\beta(\xi,\zeta)=\rho_r g + D(\xi^2+\zeta^2)^2$.  To find $u$ itself one needs to do the inverse Fourier transform, and this could, potentially, be done by the FFT.  In fact we will compute more directly with \eqref{pde}, and we will avoid the integral over time in \eqref{fouriersoln}.

Appendix \ref{app:greenthinking} illustrates the relationship of formula \eqref{fouriersoln} to the Hankel transform formulas in section \ref{sect:twolin}.

\subsection*{Implementation}  We now treat PDE \eqref{pde} numerically by discretizing using a finite difference method in time and then computing the action of $\lap^{1/2}$ and $\lap^2$ using the FFT.  Our method produces a ``whole new ball game'' numerically relative to integral formulations (the LRM method).  The resulting new method can be called a \emph{Fourier spectral collocation} method \cite{Trefethen}.

We discretize in time by the trapezoid rule---analogous to the Crank-Nicolson method for the heat equation \cite{MortonMayers}---and get an unconditionally stable $O(\Delta t^2)$ method for equation \eqref{pde}.  In particular, let $t_n=n\Delta t$ for $n=0,1,2,3,\dots$ and let $U^n(x,y)$ be our approximation of $u(x,y,t_n)$.  Equation \eqref{pde} is approximated by
\begin{align}
&\left(2\eta\,\lap^{1/2}\,U^{n+1}\right) + \frac{\Delta t}{2}(\rho_r g U^{n+1} +D\lap^2 U^{n+1}) \label{semipde}\\
&\qquad = \left(2\eta\,\lap^{1/2}\,U^{n}\right) - \frac{\Delta t}{2}(\rho_r g U^{n} +D\lap^2 U^{n}) + \Delta t\,\sigma_{zz}(x,y,t^*). \notag
\end{align}
Here either $\sigma_{zz}(x,y,t^*)=\sigma_{zz}(x,y,(n+1/2)\Delta t)$ if the load is known at the time $t^*=(n+1/2)\Delta t$ or $\sigma_{zz}(x,y,t^*)=
\frac{1}{2}\left(\sigma_{zz}(x,y,t_{n})+\sigma_{zz}(x,y,t_{n+1})\right)$ if the load is only known at the times $t_n$, $t_{n+1}$; both choices preserve $O(\Delta t^2)$ accuracy and unconditional stability.

Equation \eqref{pde} needs boundary conditions and, in fact, we assume $u(x,y,t)\to 0$, and similarly for a sufficient number of its derivatives, as $(x,y)\to \infty$.  We also assume that the support of the continuous function $\sigma_{zz}(x,y,t)$ is bounded for each $t$.  That is, we assume there is a zero boundary condition at infinity for the rigid plate and that the load is zero at sufficient distance from the origin.

With careful attention to the boundary condition at infinity, our PDE in its time-discretized form, namely equation \eqref{semipde}, can be well-approximated by its discrete Fourier transform (DFT) version.\footnote{The \emph{discrete Fourier transform} is the name of the mathematical operation; the FFT is an algorithm for computing the DFT \cite{BriggsHenson}.}  A very reasonable way to incorporate the DFT is to \emph{assume periodicity} in the spatial variables.\footnote{Other boundary conditions could be applied along the boundary of $\Omega$---e.g.~a clamped condition---but none of the easily implementable choices are obviously superior.}  For convenience we will also assume a \emph{square} region.  In fact, we assume $L$ is the half-length of a computational domain $(x,y)\in \Omega = [-L,L]\times[-L,L]$.  This domain may be substantially larger in practice than the desired region of physical interest $[-L_x,L_x]\times[-L_y,L_y]$.  We will apply periodic boundary conditions at $x,y=\pm L$ and we want $L$ to act like $\infty$ when we do this.  See figure \ref{fig:subregion}.

\begin{figure}[ht]
\widefigure{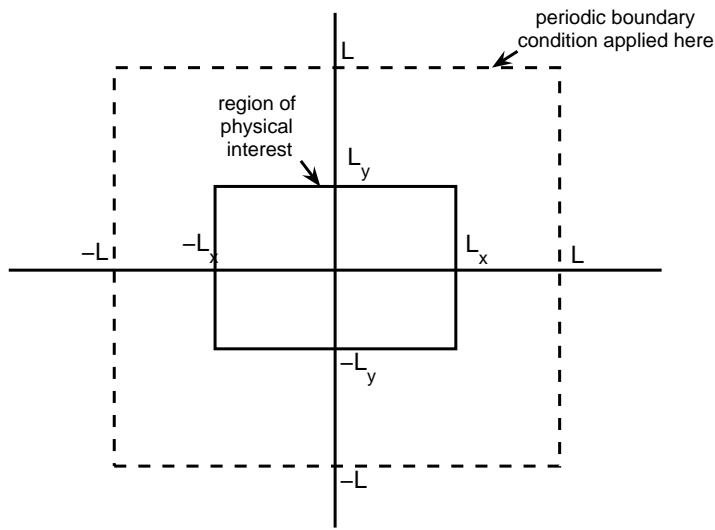}{4.4}
\vspace{-0.3in}
\caption{Fast Fourier Transform methods require periodicity.  We impose periodic boundary conditions significantly far outside the region of modelled load changes or significant deformation.} \label{fig:subregion}
\end{figure}

The PDE problem to which we apply the DFT is, therefore, Equation \eqref{semipde} on the interior of $\Omega=[-L,L]\times[-L,L]$ with periodic boundary conditions on the boundary of $\Omega$, and with the
initial condition that $U^0(x,y)$ is known on $\Omega$.

Note that from \eqref{semipde}, if $L=\infty$ then the (non-discrete)  Fourier transform $\tilde U^n =\mathcal{F}_2 U^n$ satisfies iteration
\begin{equation}\label{visciter}
\tilde U^{n+1}(\xi,\zeta) = \frac{\left[2\eta\kap - (\Delta t/2)\left(\rho_r g + D \kap^4\right)\right] \,\tilde U^n(\xi,\zeta) + \Delta t\, \tilde \sigma_{zz}(\xi,\zeta,t^*)}{2\eta\kap + (\Delta t/2)\left(\rho_r g + D \kap^4\right)}
\end{equation}
where $\kap^2=\xi^2+\zeta^2$.  This is an easily-computed iteration if one can do the (non-discrete) Fourier transform $\mathcal{F}_2$ exactly.

To use the DFT we transform the problem to a standard region $\bar \Omega=[-\pi,\pi]\times[-\pi,\pi]$.  Let $X=\pi x/L$, $Y=\pi y/L$.  Then \eqref{semipde} is equivalent to
\begin{align}
&\left(2\eta\mu\,{\bar\lap}^{1/2}\,U^{n+1}\right) + \frac{\Delta t}{2}(\rho_r g U^{n+1} +D \mu^4 {\bar\lap}^2 U^{n+1}) \label{semipdeXY}\\
&\qquad = \left(2\eta\mu\,{\bar\lap}^{1/2}\,U^{n}\right) - \frac{\Delta t}{2}(\rho_r g U^{n} +D \mu^4 {\bar\lap}^2 U^{n}) + \Delta t\,\sigma_{zz}(X,Y,t^*)  \notag
\end{align}
on $\bar\Omega$, where $U^n=U^n(X,Y)$, $\mu=\pi/L$, and $\bar\lap=-\left(\partial^2/\partial X^2+\partial^2/\partial Y^2\right)$.

Let $N$ be an integer; typically $N$ is a power of $2$ for efficiency in the FFT.  Let $h=2\pi/N$ and let $X_j=-\pi+jh$, $Y_k=-\pi+kh$, $j,k=1,\dots,N$.  On $\bar\Omega$ we use the DFT, as normalized by \citeasnoun{Trefethen}, in variables $X,Y$.  If $f(X,Y)$ is some function on $\bar\Omega$ with grid values $f_{jk}=f(X_j,Y_k)$ then the DFT (forward, inverse) pair is
\begin{equation}\label{DFT}
\hat f_{pq} = h^2 \sum_{j,k=1}^N e^{-i(p X_j + q Y_k)} f_{jk}, \qquad 
f_{jk} = \frac{1}{(2\pi)^2} \sum_{p,q=-N/2+1}^{N/2} e^{i(p X_j + q Y_k)} \hat f_{pq}.
\end{equation}

Let
\begin{equation}\label{bandlimited}
\bar f(X,Y) = \frac{1}{(2\pi)^2} \sum_{p,q=-N/2+1}^{N/2} e^{i(p X + q Y)} \hat f_{pq}
\end{equation}
be the ``band-limited trigonometric interpolant'' of $f(X,Y)$ \cite{Trefethen}; note the relation to the inverse DFT \eqref{DFT}.  We see that
    $$\bar\lap \bar f(X,Y) = \frac{1}{(2\pi)^2} \sum_{p,q=-N/2+1}^{N/2} (p^2+q^2) e^{i(p X + q Y)} \hat f_{pq}.$$
That is, the Laplacian $\bar\lap$ on $\bar\Omega$ corresponds to multiplying the $p,q$ mode by $(p^2+q^2)$.  Thereby $\bar\lap^{1/2}$ and $\bar\lap^2$ are also defined, respectively, by multiplication by $(p^2+q^2)^{1/2}$ and $(p^2+q^2)^2$.

It now follows that \eqref{semipdeXY} is very easy to compute if we approximate $U^n(X,Y)$ by its band-limited interpolant $\bar U^n(X,Y)$ and compute the action of the powers of the Laplacian by the multipliers above.  This describes a \emph{Fourier spectral collocation method}.  That is, one time-step in solving PDE \eqref{pde} by our method is the sequence
\renewcommand{\labelenumi}{(\emph{\roman{enumi}})}\begin{enumerate}
\item compute the DFT $\hat U^n_{pq}$ by FFT from values $U^n_{jk} \approx U^n(X_j,Y_k) = U^n(x_j,y_k)$; also compute the DFT of the load $(\hat\sigma_{zz})_{pq}$ at $t=t^*$ from values $\sigma_{zz}(x_j,y_k,t^*)$,
\item compute
\begin{equation}\label{DFTpde}
\hat U^{n+1}_{pq} = \frac{\left[2 \eta\mu (p^2+q^2)^{1/2} - (\Delta t/2)\left(\rho_r g + D \mu^4 (p^2+q^2)^2\right)\right] \hat U^n_{pq} + \Delta t\,\left(\hat \sigma_{zz}\right)_{pq}}{2 \eta\mu (p^2+q^2)^{1/2} + (\Delta t/2)\left(\rho_r g + D \mu^4 (p^2+q^2)^2\right)},
\end{equation}
where $\mu=\pi/L$ and
\item undo the DFT (i.e.~do the inverse FFT and make sure the result is real) to get $U^{n+1}_{jk}\approx U^{n+1}(X_j,Y_k) = U^{n+1}(x_j,y_k)$.
\end{enumerate}
Compare equation \eqref{DFTpde} to equation \eqref{visciter} which applies for the non-discrete Fourier transform $\tilde U^n$.

Full \Matlab implementations of the methods in this report are given in Appendix \ref{app:matlab}.  Only a few lines of \Matlab are needed to implement the core sequence above, however:
\vspace{0.1in}
\begin{quote}
\noindent\texttt{for n=0:M-1}

[computations using current displacement \texttt{uun} $=U^n(x_j,y_k)$]

\texttt{uun=uun1;}

[get \texttt{H} $=H(x_j,y_k,t^*)$]

\texttt{sszz=-rhoi*g*H;}

\texttt{frhs=right.*fft2(uun) + fft2(dt*sszz);}

\texttt{uun1=real(ifft2( frhs./left ));}

\noindent\texttt{end}
\end{quote}
\vspace{0.1in}
Here ``\texttt{right}'' and ``\texttt{left}'' are pre-computed grid values of the expressions $2 \eta\mu k - (\Delta t/2)B$ and $2 \eta\mu k + (\Delta t/2)B$ which appear in \eqref{DFTpde}, respectively, where $k^2=p^2+q^2$ and $B=\rho_r g + D \mu^4 k^4$.

We call the iteration (\emph{i}), (\emph{ii}), (\emph{iii}) the ``PDE method'' in contrast to the load response matrix method (``LRM method'') of the previous section.  With standard estimates on the speed of the FFT when $N_x$, $N_y$ are powers of two \cite{BriggsHenson}, the ``PDE method'' requires
\begin{equation}\label{PDEest}
O(N_x N_y (\log_2 N_x) (\log_2 N_y))
\end{equation}
scalar operations to update the vertical displacement.  This compares directly to equations \eqref{LRMest} and \eqref{LRMestwFFT} for the ``LRM method.''  In particular, relative to the $O(N_t N_x N_y (\log_2 N_x) (\log_2 N_y))$ estimate \eqref{LRMestwFFT} for the ``LRM method'' using the FFT for convolution we note a factor of $N_t$ less work.  Furthermore, the ``PDE method'' avoids the entire stage of computing the LRM integrals, which turn out to be quite expensive, though totally uninteresting, computations.

Relative to the ``LRM method'' \emph{without} the FFT, using representative values of $N_t=25$ and $N_x=N_y=80$, and assuming that the constants in the ``big $O$ notation'' are about the same, we get a speed up of about
    $$(25 \cdot 80^2 \cdot 80^2)/(80^2 (\log_2 80)^2) \approx 4 \times 10^3.$$
This ratio is roughly what we observe in practice.  For instance, on the same computer we compared Fortran 77 codes running the ``LRM method'' sans FFT using $N_t=101$ and $N_x=N_y=31$ with \Matlab codes (Appendix \ref{app:matlab}) implementing the ``PDE method'' using $N_x=N_y=32$.  We used $\Delta t=500$ years and bed deformations were computed for 50k years in both cases.  The former method took about 10 \emph{hours} while the later took about 4 \emph{seconds} for an observed speedup of about $9 \times 10^3$.

\section{Discussion}\label{sect:discuss}

In the next section we describe the results of computations with the ``PDE method.''  It is appropriate, however, to first directly address the apparently new feature in PDE \eqref{pde},
    $$\ppt{}\left(2\eta\,\lap^{1/2}\,u\right) + \rho_r g u +D\lap^2 u = \sigma_{zz},$$
namely the viscosity expression ``$\partial/\partial t \left(2\eta \lap^{1/2} u\right)$.''

Within the ice sheet modeling community there is a simplified existing \emph{standard model} for a flat ``elastic plate (lithosphere) that overlies a viscous asthenosphere'' \cite{Greve2001,HagdornThesis,ZweckHuybrechts}.  Comparison to this model illuminates the significance of the viscosity expression in \eqref{pde}.  The standard model consists of two equations
\begin{gather}
\rho_r g w + D \lap^2 w = \sigma_{zz}, \label{notionalplate} \\
\ppt{u^s} = - \frac{u^s - u_0- w}{\tau}, \label{notionaldecay}
\end{gather}
where $u^s(x,y,t)$ is the vertical displacement of the earth's surface, $u_0(x,y)$ is a hypothesized unloaded displacement state, and $w(x,y,t)$ is the position of a notional elastic plate which is in equilibrium with the current load $\sigma_{zz}(x,y,t)$.  The essential viscous constant for the standard model is a characteristic time scale $\tau$ of relaxation, chosen, for example, as $3000$ years by \citeasnoun{Greve2001} and \citeasnoun{ZweckHuybrechts}.  The relaxation time $\tau$ is indirectly related to the asthenosphere viscosity $\eta$; more on this below.

We can now calculate an illuminating comparison by hand.  Suppose that at time $t=0$ all load is removed but that the vertical displacement is a $y$-independent sinusoidal mode with spatial frequency $k$:
\begin{equation}\label{initmode}
u(x,y,t\!=\!0) = u^s(x,y,t\!=\!0) = A_0 \exp(ik\pi x/L).
\end{equation}
Here $A_0$ is the initial amplitude and $L$ is a characteristic length scale.  We ask: how does such a mode decay in the two models?

Note that the $q$th power of the Laplacian act on this mode as follows:
    $$\lap^q \exp(ik\pi x/L) = (k^2\pi^2/L^2)^q \exp(ik\pi x/L).$$
Thus model \eqref{pde} has solution $u(x,y,t) = A(t) \exp(ik\pi x/L)$ where
    $$2\eta (|k|\pi/L) \dot A + \rho_r g A + D(k^4\pi^4/L^4) A = 0.$$
That is, in model \eqref{pde} the amplitude of the $k$ mode satisfies
\begin{equation}\label{odenew}
\dot A = - \frac{\rho_r g L^4 + D k^4 \pi^4}{2\eta L^3 |k| \pi} A, \qquad A(t\!=\!0) = A_0,
\end{equation}
so a mode with frequency $k$ decays a rate that depends upon $k$.  By contrast, in model \eqref{notionalplate}, \eqref{notionaldecay} with $u_0=0$ the same mode evolves by
\begin{equation}\label{odestandard}
\dot A = -\frac{1}{\tau} A, \qquad A(t\!=\!0) = A_0
\end{equation}
because $w=0$ as the load has been removed; equation \eqref{notionaldecay} reduces to $\partial u^s/\partial t = - \tau^{-1} u^s$.  Here we see that all modes decay at a rate independent of the frequency.

But it is clearly the case that a viscous asthenosphere will make elastic plate modes decay at different rates depending on the frequency.  Indeed, \citeasnoun{Greve} identifies the failure of the standard model \eqref{notionalplate}, \eqref{notionaldecay} to have frequency dependent relaxation times as a deficiency of that model relative to full spherical self-gravitating models.  Comparing equation \eqref{odenew} to \eqref{odestandard} we are motivated to plot the function
\begin{equation}\label{tauofk}
\tau(k)=\frac{2\eta L^3 |k| \pi}{\rho_r g L^4 + D k^4 \pi^4},
\end{equation}
which has units of time.  Supposing $L=2000$ km and that $\rho_r,g,D,\eta$ have the values given in section \ref{sect:twolin}, we plot $\tau(k)$ in figure \ref{fig:tauofk}.  We see that the standard choice $\tau=3000$ a in \eqref{odestandard} corresponds to frequencies $k\approx 1$ and $k\approx 10$ in \eqref{odenew}, but that no constant relaxation time is representative of the actual relaxation spectrum.

\begin{figure}[ht]
\widefigure{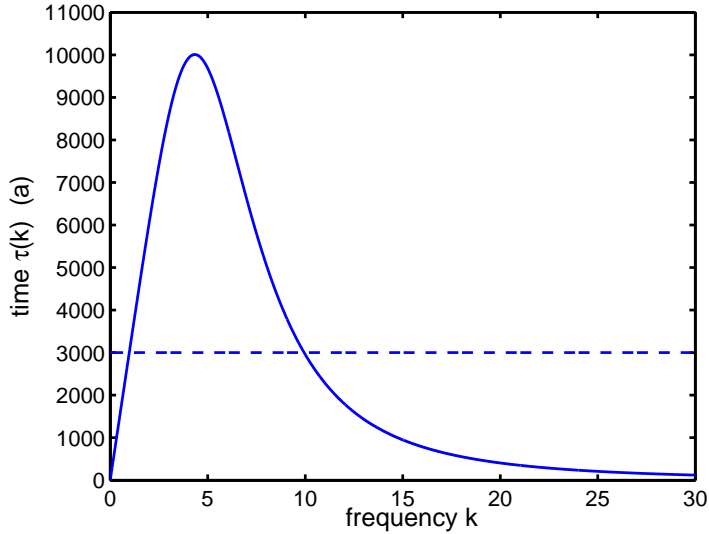}{4.0}
\caption{Frequency dependent relaxation time $\tau(k)$ (solid) for mode $k$ in equation \eqref{initmode}.  The value $\tau=3000$ a (dashed) for the standard model is a reasonable constant value, but no constant provides a good fit.} \label{fig:tauofk}
\end{figure}

Furthermore we see in formula \eqref{tauofk} and figure \ref{fig:tauofk} that for small $k$ (e.g.~$k\lesssim 3$) the relaxation time $\tau(k)$ is proportional to $k$.  This behavior is identified by \cite{KlemannWolf}---see \citeasnoun[section 5]{Greve2001}---as correct for the most significant mode in a spherical, self-gravitating viscoelastic earth model.

The justification for model \eqref{notionalplate}, \eqref{notionaldecay} is its computability, of course.  Indeed, the computation of the elliptic PDE \eqref{notionalplate} is standard in all numerical paradigms (finite difference, finite element, spectral).  At least on a rectangular spatial grid, however, the time-semi-discretization \eqref{semipde} of equation \eqref{pde} is just as computable as \eqref{notionalplate}.  In particular, if an ice sheet simulation is performed on a rectangular grid using a finite difference or finite element method for the ice dynamics then equation \eqref{pde} can be easily computed by the Fourier spectral collocation method of the previous section.

Now we come to another reason to prefer equation \eqref{pde} as a model for earth deformation in the context of ice sheet modeling.  Let us suppose that at the current time the ice thickness $H_0$ (and possibly water depth, giving an effective ice thickness) in a region of interest has been well-measured.  Let us suppose that a reasonably detailed map of current uplift rate $\dot u_0=\partial u/\partial t$ is also known.  This is a realistic supposition given given the state of observational geophysics circa 2006 because uplift can be well-constrained by GPS measurements (Larsen and others 2005)\nocite{Larsenetal} when bedrock is exposed.  Alternatively a spherical viscoelastic earth model of more-or-less arbitrary sophistication and computational expense might generate a trusted current uplift map \cite{IvinsJames2005}.  In either case we can then use \eqref{pde} to determine the initial condition for the earth deformation part of an ice sheet simulation without requiring further reference to an assumed past load history; compare the integration over load history scheme used in \cite{LingleClark}.  In fact, by \eqref{pde} we may solve
\begin{equation}\label{prebent}
\rho_r g u_0 + D \lap^2 u_0 = \rho_i g H_0 - 2\eta \lap^{1/2} \dot u_0
\end{equation}
for $u_0$ to get the starting displacement.  In other words we ask for the ``pre-bent'' position of the elastic plate in the half space model which  accounts for the current uplift rate using the current load (i.e.~current effective ice thickness).  Solving \eqref{prebent} numerically is no harder than, in fact it amounts to, one step of the numerical method already described.  If the load does not change in the simulation, an uninteresting case for ice sheets of course, then the elastic plate overlying the viscous half space will start at the current time with the current uplift but will then relax to the state satisfying equation \eqref{equilbouyant} for equilibrium with the load and bouyant force, and there will no more uplift.  Note that the presence of bed topography is completely irrelevant here because of the linear nature of the model; see comments in \cite{LingleClark} to the effect that bed topography represents a irregular ``thin veneer of zero strength'' atop an elastic plate lithosphere of (significant) flexural rigidity $D$.

The mechanism described in the previous paragraph is, we believe, a more principled replacement for the hypothesized ``unloaded surface elevation'' $u_0(x,y)$ used in the standard model \eqref{notionalplate}, \eqref{notionaldecay}.  Use of that standard model seems to require an assumption of present day isostatic equilibrium with the current load \cite{ZweckHuybrechts} or other artificial assumption which is in conflict with observed spatially-varying current uplift.

An entirely different class of viscoelastic earth models exists in the literature, of course.  These are the layered, spherical, self-graviating viscoelastic models which typically descend from the work of \citeasnoun{Cathles} and/or \citeasnoun{Peltier}.  The numerical implementation of these models typically involves computing a high degree spherical harmonic expansion of the strain field for the entire three-dimensional geoid.  The traditional difficulty with these models is their computational cost \cite{Fastook,Greve2001}.  Furthermore there is only modest benefit because results from the standard model \eqref{notionalplate}, \eqref{notionaldecay} above, in particular, are regarded by the ice sheet modeling community as reasonably close to those from the spherical models \cite{Greve2001}.  Equation \eqref{pde} is promising because it is just as computationally inexpensive as the standard model but incorporates at least one important feature of the spherical models, namely frequency dependent relaxation times.

Speaking mathematically, an interesting additional possibility exists within the same class of computationally inexpensive ``PDE methods.''  Namely, one should be able to modify equation \eqref{pde} to take spherical effects into account.  In particular, equation \eqref{pde} would be computable at essentially the same speed if it were replaced by a non-constant coefficient version, for instance
\begin{equation}\label{possibleform}
\ppt{}\left(2 \eta(x,y) \lap^{1/2} u\right) + \rho_r g \alpha(x,y) u + D(x,y) \lap\left(\beta(x,y) \lap u\right) = \sigma_{zz}.
\end{equation}
We do not currently know that a set of non-constant coefficients $\eta,D,\alpha,\beta$ exist which correctly account for spherical geometry.  It seems, however, that classical continuum mechanics and differential geometry must produce such a form because the standard model for a spherical self-gravitating earth is a \emph{linear} model.  That is, it has linear response to load.  Abstractly, this linearity is all that is necessary to make it  inevitable\footnote{See the discussion of distributions in \cite{ReedSimon}, for instance.} that a two (spatial) dimension, non-constant coefficient linear equation for the vertical displacement of the earth's surface must exist for each patch of the earth's surface.  It may well involve additional pseudo-differential operators not present in the putative form \eqref{possibleform}, however.

\section{Numerical results} \label{sect:results}

\subsection*{A ``tweak'' to the procedure}  It turns out that a small error can be avoided if the ``PDE method'' is modified slightly.  In fact, in verifying the ``PDE method'' below, using an exact integral formula for a disc load, we observed that there was a uniform a error of several meters.  This uniform error decayed slowly as the distance at which the periodic boundary condition was applied went to infinity.

The following simple modification eliminates this error.  Using the notation of section \ref{sect:PDEmethod}, consider $U^n(x_j,y_k)$ on the grid at timestep $n$.  Let $\bar U^n_L$ be the average value of $U^n$ along the boundary of the computational domain $\Omega=[-L,L]\times[-L,L]$; recall $\Omega$ is typically larger than the physical region $[-L_x,L_x]\times[-L_y,L_y]$.  Let $u_{H_0,R_0}^\infty(r)$ be the vertical displacement at distance $r$ from the center of an ice disc load of thickness $H_0$ and radius $R_0$ of an elastic plate in equilibrium with the bouyant restoring force.  That is, let $u_{H_0,R_0}^\infty(r)$ be the value from formula \eqref{equildisc} in Appendix \ref{app:exactdisc}.   Choose the values $H_0,R_0$ so that the volume $\pi R_0^2 H_0$ of the disc load matches the current (timestep $n$) load volume, or rather its ice equivalent volume if appropriate.

Our ``tweak'' replaces the solution $U^n(x_j,y_k)$ at each timestep $n$ with values to which a constant shift has been applied:
\begin{equation}\label{tweak}
U^{n,\circ}(x_j,y_k)=U^n(x_j,y_k)-\bar U^n_L+u_{H_0,R_0}^\infty(L).
\end{equation}
That is, we want the ``far-field value'' produced by the original ``PDE method'' to be thrown out and replaced by the equilibrium plate value with an equivalent disc load.  Though the volume of the equivalent disc is determined by the current load, one obviously has some freedom in choosing its thickness and radius.  We presume that an effort is made to approximate the aspect ratio of the actual load, but close matching is not essential.  In fact the Green's function value would work reasonably well, too.

\subsection*{Verification}  The first concern regarding computations with our earth deformation model is \emph{verification}.  In particular, we want to know if numerical results from the ``PDE method'', with the just-mentioned ``tweak,'' are close to highly-accurate solutions of the continuum equation \eqref{pde}.  In seeking such solutions we inevitably come to disc loads.  Appendix \ref{app:exactdisc} addresses this case by the Hankel transform.  It yields equation \eqref{discint}, an integral formula for the time-dependent radially-symmetric deflection $u^V(r,t)$ resulting from the application (at time zero) of a disc load; see figure \ref{fig:discload}.  The integral must be computed numerically, but numerical quadrature is an approximation completely independent of the cartesian-grid- and FFT-based ``PDE method.''  We use integral formula \eqref{discint} as an ``exact'' solution, believing the accuracy of our numerical integration of \eqref{discint} to exceed that of the ``PDE method'' for any achievable grid.

Our \Matlab implementation of the ``PDE method'' is the function \mtt{fastearth.m} listed in Appendix \ref{app:matlab}.  Also listed are implementations of the LRM method for the spherical elastic earth model (\mtt{geforconv.m}) and the numerical integration of equation \eqref{discint} (\mtt{viscdisc.m}).

Let us define a particular numerical experiment.  Suppose we use the parameters specified in section \ref{sect:twolin}: $D=5.0\times 10^{24}$ N m, $\rho_r=3300$ $\text{kg}\,\text{m}^{-3}$, and $\eta=10^{21}$ $\text{Pa}\,\text{s}$.  Suppose the disc of ice (density $\rho_i=910$ $\text{kg}\,\text{m}^{-3}$) has radius $1000$ km and thickness $1000$ m.  We seek the deflection on a square region $R$, centered on the disc load, of side length $4000$ km ($L_x=L_y=2000$ km).  Suppose that at time zero the deflection is identically zero, and suppose that the load is applied at that time.  We calculate the  deflection at $t=20k$ years as computed by the ``PDE method'' and by formula \eqref{discint}.

There are \emph{three} numerical parameters of importance for the ``PDE method'': \begin{itemize}
\item $N$, the number of grid points in each direction,
\item $Z$, the factor by which the computational domain $\Omega=[-L,L]\times[-L,L]$ is larger than the physical domain $[-L_x,L_x]\times[-L_y,L_y]$; here $L_x=L_y$, and
\item $\Delta t$, the time step used in approximating the time derivative which occurs in equation \eqref{pde}.\end{itemize}
Our verification involves showing that as these parameters go to their continuum limits ($N\to\infty$, $Z\to\infty$, $\Delta t\to 0$) we approach the ``exact'' solution \eqref{discint}.

For verification we first fixed $\Delta t=100$ years and considered the effect of grid refinement and of changes to the distance at which the periodic boundary condition was applied.  Regarding grid refinement we considered $N$ ranging over powers of two from $2^3=8$ to $2^8=256$.  Regarding the distance to the periodic boundary condition we imposed the periodic boundary condition at $|x|,|y|=L$, where $L=Z \,L_x=Z\,L_y$, and we used $Z=1,2,4,8$, but we quickly discovered that with the above-mentioned ``tweak'' any value $Z\ge 2$ works fine; not shown.

The result for \emph{maximum error} under grid refinement is shown in figure \ref{fig:maxerr}.  This maximum error is not the only reasonable measure.  As shown in figure \ref{fig:errmap}, for fine grids errors greater than one meter are highly localized to certain points just at the edge of the disc load.  Note realistic ice loads do not have margins as sharp as this disc load.  For these reasons among others it is reasonable to consider average errors, and we see in figure \ref{fig:averr} that the average errors are less than $20$ cm for $N=256$, or roughly $0.07\%$ of the compensation depth.

\begin{figure}[ht]
\widefigure{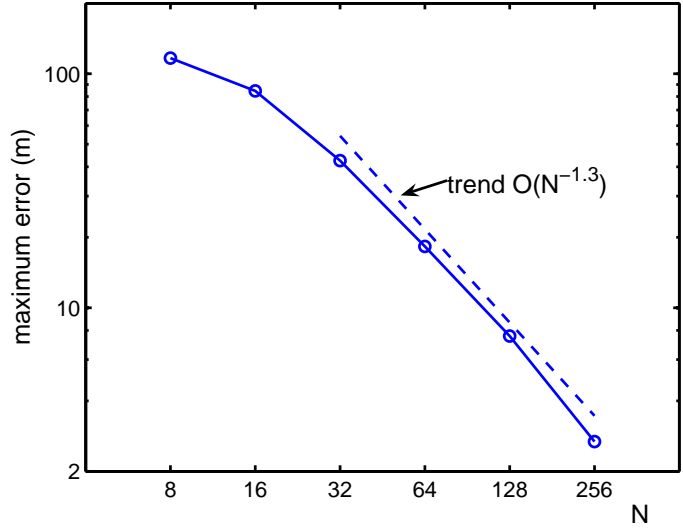}{4.0}
\caption{Maximum error made by the ``PDE method'' relative to the Hankel transform integral \eqref{discint}.  Grid refinement (increasing $N$) reduces the max error to below $3$ m when $N=256$.  Here $Z=2$ and $\Delta t=100$ a.} \label{fig:maxerr}
\end{figure}

\begin{figure}[ht]
\widefigure{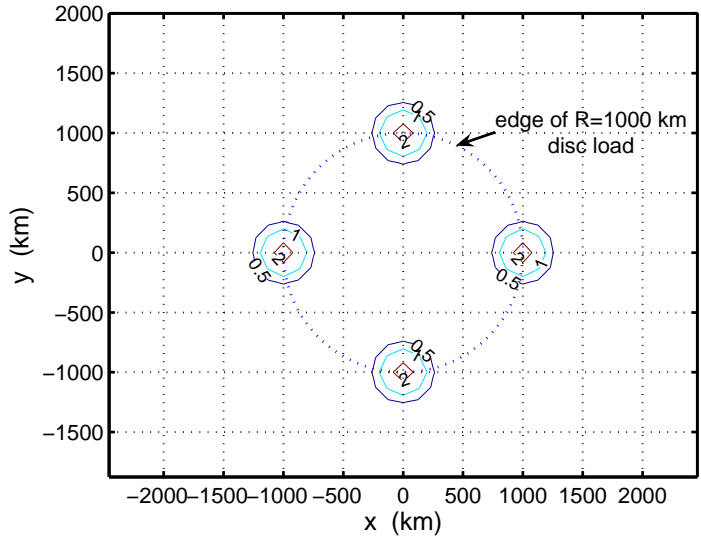}{4.0}
\caption{Spatial distribution of the error when $N=256$, $Z=2$, and $\Delta t=100$ a.  Contours of the error $|(\text{PDE method})-(\text{equation \eqref{discint}})|$ at 0.5, 1, 2 m.  The error is concentrated where the edge of the disc load meets the coordinate axes.} \label{fig:errmap}
\end{figure}

\begin{figure}[ht]
\widefigure{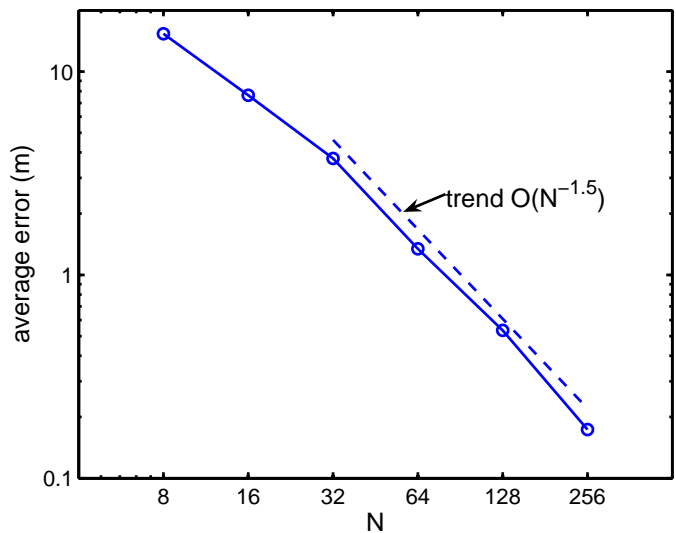}{4.0}
\caption{Same as figure \ref{fig:maxerr} but now \emph{average} error.  Grid refinement (increasing $N$) reduces the average error to below $20$ cm when $N=256$.} \label{fig:averr}
\end{figure}

Next we compare the effects of spatial grid refinement, increasing $N$, to reduction of time stepsize $\Delta t$ on the error.  See figure \ref{fig:averrdt}.  We see that any value of $\Delta t$ less than $500$ years is fine; this is great news for ice sheet simulation.  It is not, however, surprising because of the relative timescales of ice versus asthenosphere flow.

\begin{figure}[ht]
\widefigure{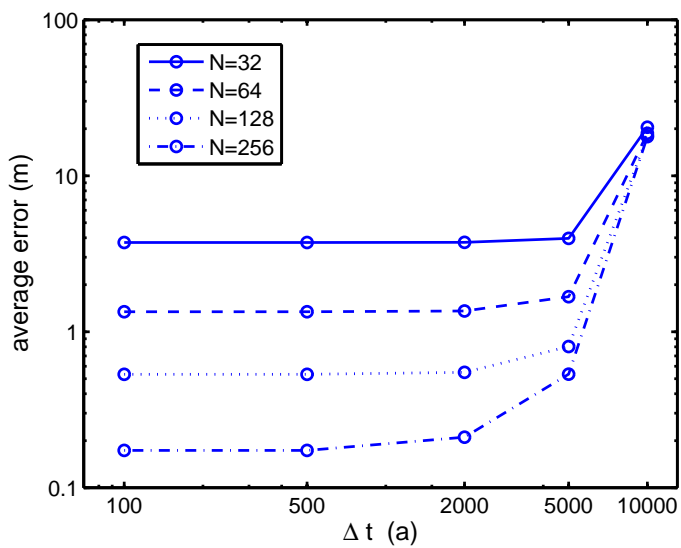}{4.0}
\caption{Average error as in figure \ref{fig:averr} but with $\Delta t$ varying, and  for several values of $N$.  There is no need for $\Delta t < 500$ a.  Spatial grid refinement (increasing $N$) is more important to reducing error than is temporal refinement (decreasing $\Delta t$).} \label{fig:averrdt}
\end{figure}

\subsection*{Ice sheet modeling}  Earth deformation used in the context of ice sheet modeling is our actual interest.  Earth deformation is obviously effected by ice sheet flow---the load moves around.  Conversely, as the bed deforms the surface slope of the ice sheet changes and this effects flow.  There is non-trivial coupling.

The reasonable simplest ice sheet model is the isothermal model with Glen rheology \cite{Paterson,Nye00}.  Let $h(x,y,t)$ be the surface elevation of the ice and let $H(x,y,t)$ be the ice thickness.  The frozen-base isothermal ice sheet equation is the single nonlinear diffusive partial differential equation
\begin{equation}\label{isie}
\ppt{H} = M + \Div\left(\Gamma\, H^{n+2} |\grad h|^{n-1} \grad h\right)
\end{equation}
where $n$ is the Glen exponent, e.g.~$n=3$, and $\Gamma$ is a constant (typically $\Gamma=2 (\rho_i g)^n A_0/(n+2)$ where $A_0$ is a softness parameter).  If $b(x,y,t)$ is the ice sheet bed elevation---a slight change of notation from the rest of the paper---then of course $h=b+H$.

As we now show, exact similarity solutions to this equation which incorporate simple isostasy (Nye 2000, Bueler et al.~2005; compare Halfar 1983)\nocite{Halfar83,Nye00,BLKCB} provide a very nice tool to examine coupling to the earth model.  They help illuminate the differences among earth models.    By ``simple isostasy'' we mean the rule which specifies
\begin{equation}\label{simple}
b=-f H
\end{equation}
where $f$ is a fixed fraction of the ice thickness \cite{Nye00}; we will let $f=\rho_i/\rho_r=0.27576$ in our numerical experiments.  Since $h=b+H$, if equation \eqref{simple} applies then $h=(1-f)H$.

We will compare numerical results for three coupled ice sheet flow/earth deformation models:
\newcommand{\simple}{\textsc{Simple}\xspace}
\newcommand{\standard}{\textsc{Standard}\xspace}
\newcommand{\LC}{\textsc{Lingle\&Clark}\xspace}
    $$\begin{matrix}
    \text{\simple:} & \text{equations \eqref{isie} and \eqref{simple},} \\
    \text{\standard:} & \text{equations \eqref{isie}, \eqref{notionalplate}, \eqref{notionaldecay}, and $b=u^s$,} \\
    \text{\LC:} & \text{equations \eqref{isie}, \eqref{GEuse}, \eqref{pde}, and $b=u^E+u^V$.} \\
\end{matrix}$$
Figure \ref{fig:profilecompare} shows a result of a coupled simulation for these three models.  A detail near the margin is shown in figure \ref{fig:margindetail}.

\begin{figure}[ht]
\centerline{\widefigure{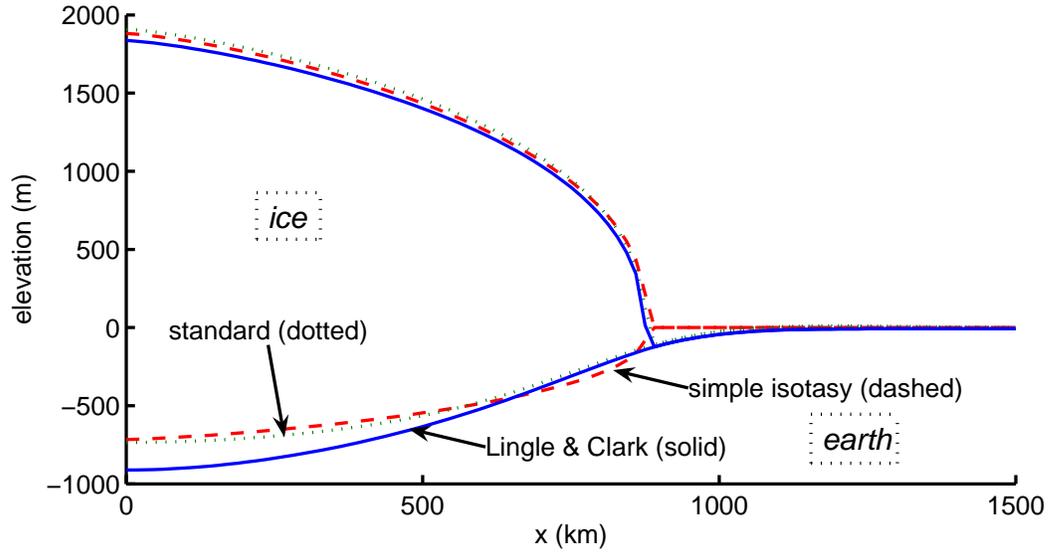}{6.0}}
\caption{Ice sheet on deforming bed, at time $60$k years, from three earth models \simple, \standard, and \LC.  View of gridded numerical values ($N_x=N_y=192$) along the positive $x$-axis of the grid.} \label{fig:profilecompare}
\end{figure}

\begin{figure}[ht]
\widefigure{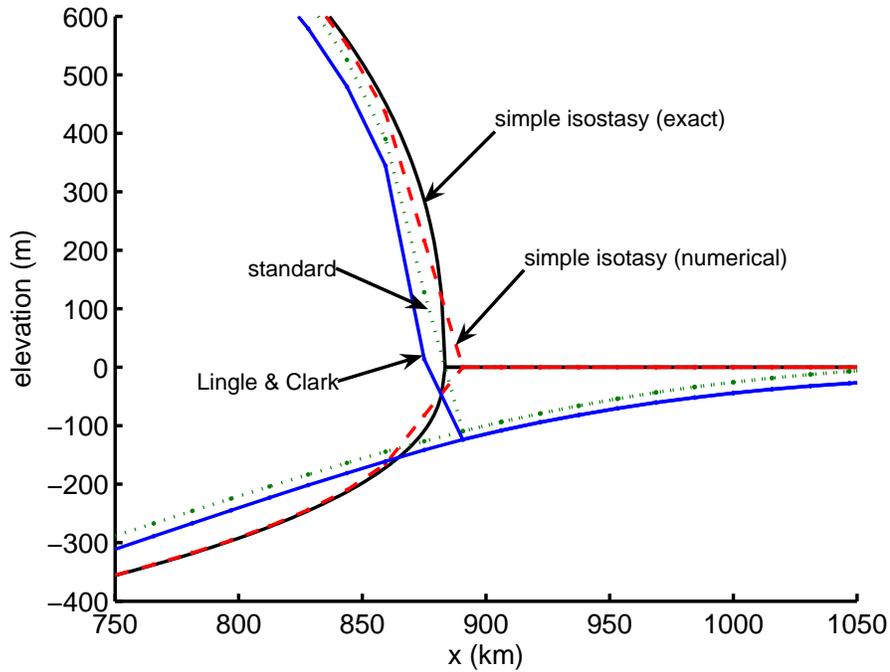}{5.0}
\caption{See figure \ref{fig:profilecompare}; detail near the grounded margin.  Exact similarity solution to the simple isostasy model is added (solid).} \label{fig:margindetail}
\end{figure}

In fact, the result shown in figure \ref{fig:profilecompare} came from starting with $H=0$ and $b=0$ at $t=0$ and using an accumulation history corresponding to the simlarity solution illustrated by figure \ref{fig:simviews}.  That is, the accumulation $M(x,y,t)$ comes from equations (9) and (10) in \Bueleretal, using $f=910/3300$, $\lam=5$, $\alpha=-1$, $\beta=2$, $H_0=3600$ m, $R_0=750$ km, $\Gamma=9.0177 \times 10^{-13}$ $\text{m}^{-3}$ $\text{s}^{-1}$, and with the additional statement $M=5 t^{-1} H_\lam$.  Note $t_0=40034$ years.  In addition, at time $t=t_0$ the accumulation is turned off and so for $t>t_0$ the exact behavior of the solution to \simple is a Halfar-type \cite{Halfar83} accumulation-free solution.  Thus the accumulation history is from a similarity solution to equation \eqref{isie}, incorporating simple isostasy, which grows from zero at $t=0$ to maximum height at $t_0=40034$ years and spreads out from then on, with no loss of volume.

\begin{figure}[ht]
\centerline{\widefigure{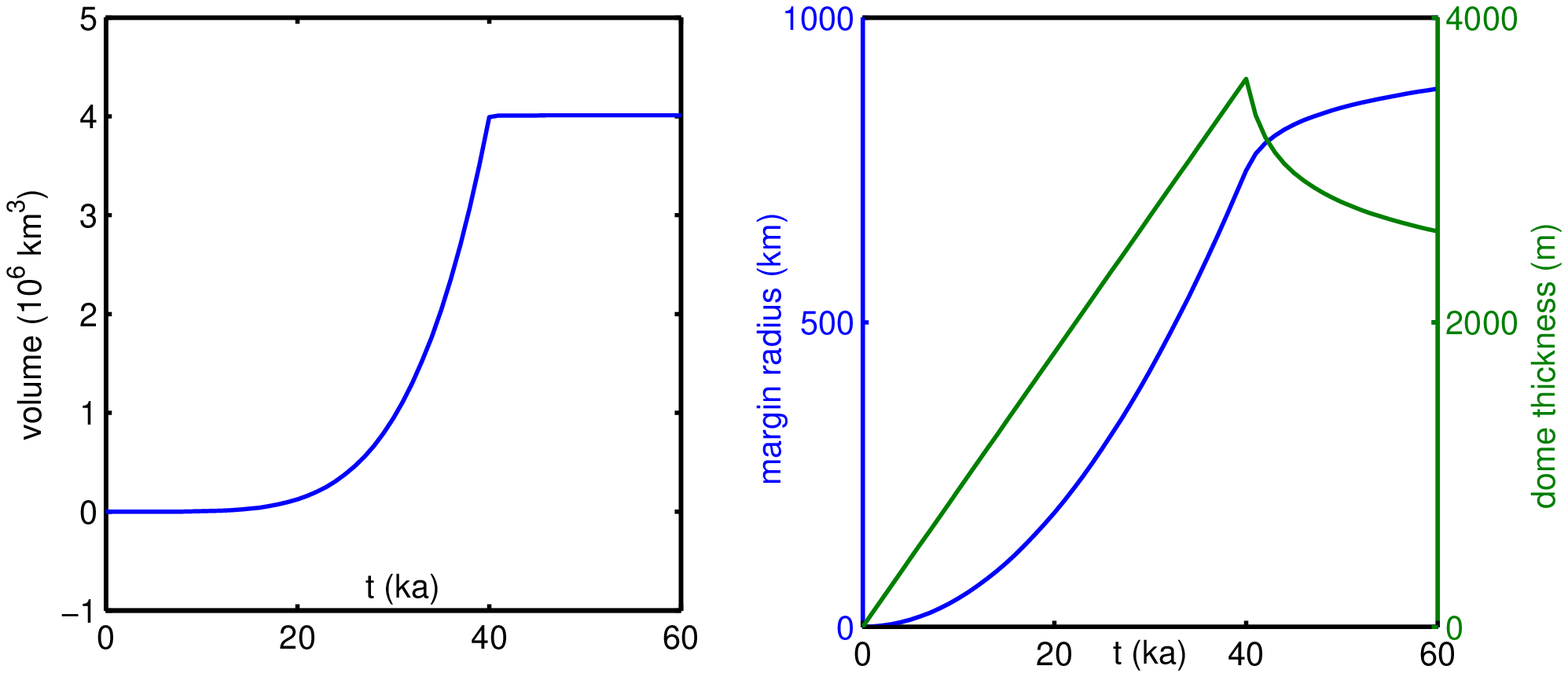}{6.0}}
\caption{Views of a similarity solution to equation \eqref{isie}.  Left: volume over time.  Right: margin radius (solid) and dome height (dashed) over time.  At time $t_0=40034$ a the solution switches from growing ($\lam=5$) to Halfar ($\lam=0$), in both cases with simple isostasy ($f=910/3300$).} \label{fig:simviews}
\end{figure}

The importance of such a similarity solution is that it forms an \emph{exact} continuum solution to the \textsc{Simple} model.  Therefore we can answer with some precision the question ``\emph{how do differences resulting from coupling to various earth deformation models compare to the numerical errors which occur in ice sheet modeling?}''  This is an important question.  If numerical ice sheet errors demonstrably exceed the earth model differences then we should be skeptical of any expenditure of effort in the earth modeling direction.  Conversely, even if the differences among earth models are significant, one should report these differences relative to the actual magnitude of numerical ice modeling errors.

Figure \ref{fig:profilecompare} indeed suggests differences among the coupled models.  We ran each model to final time $t=60k$ a.  As shown in figure \ref{fig:commonvol}, however, all of the models produce the same volume at the final time, and indeed at all times; this follows from using the same finite difference approach for the ice flow (as described in \Bueleretal) and, of course, the same accumulation history $M(x,y,t)$.  So the differences can be described by the distributions of ice thickness.  In figure \ref{fig:thickdiffs} we show the maximum and average of the pairwise absolute thickness differences $|H_{\text{\simple}}-H_{\text{\standard}}|$, etc.  (The average differences are over the $H>0$ grid points under \simple.)  We see average thickness differences greater than $10$ m between each pair.  We see that the greatest pairwise difference is between \simple and \LC; compare figure \ref{fig:profilecompare}.

\begin{figure}[ht]
\widefigure{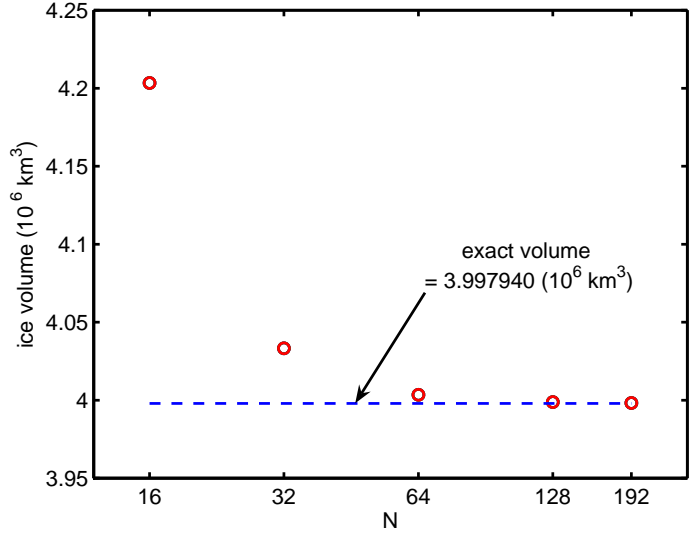}{4.0}
\caption{All models have identical convergence of numerical volume at $t=60k$ a; they share the same accumulation history.} \label{fig:commonvol}
\end{figure}

\begin{figure}[ht]
\widefigure{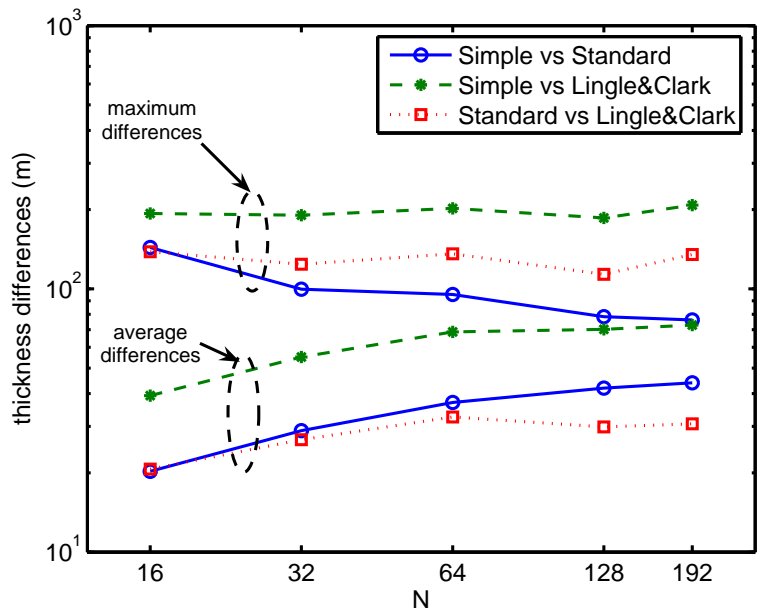}{4.5}
\caption{Maximum and average \emph{ice thickness} differences in pairwise comparison.} \label{fig:thickdiffs}
\end{figure}

We see a similar picture for bed elevation differences, with \simple versus \standard showing somewhat smaller differences, and the comparison \simple versus \LC again being largest.

\begin{figure}[ht]
\widefigure{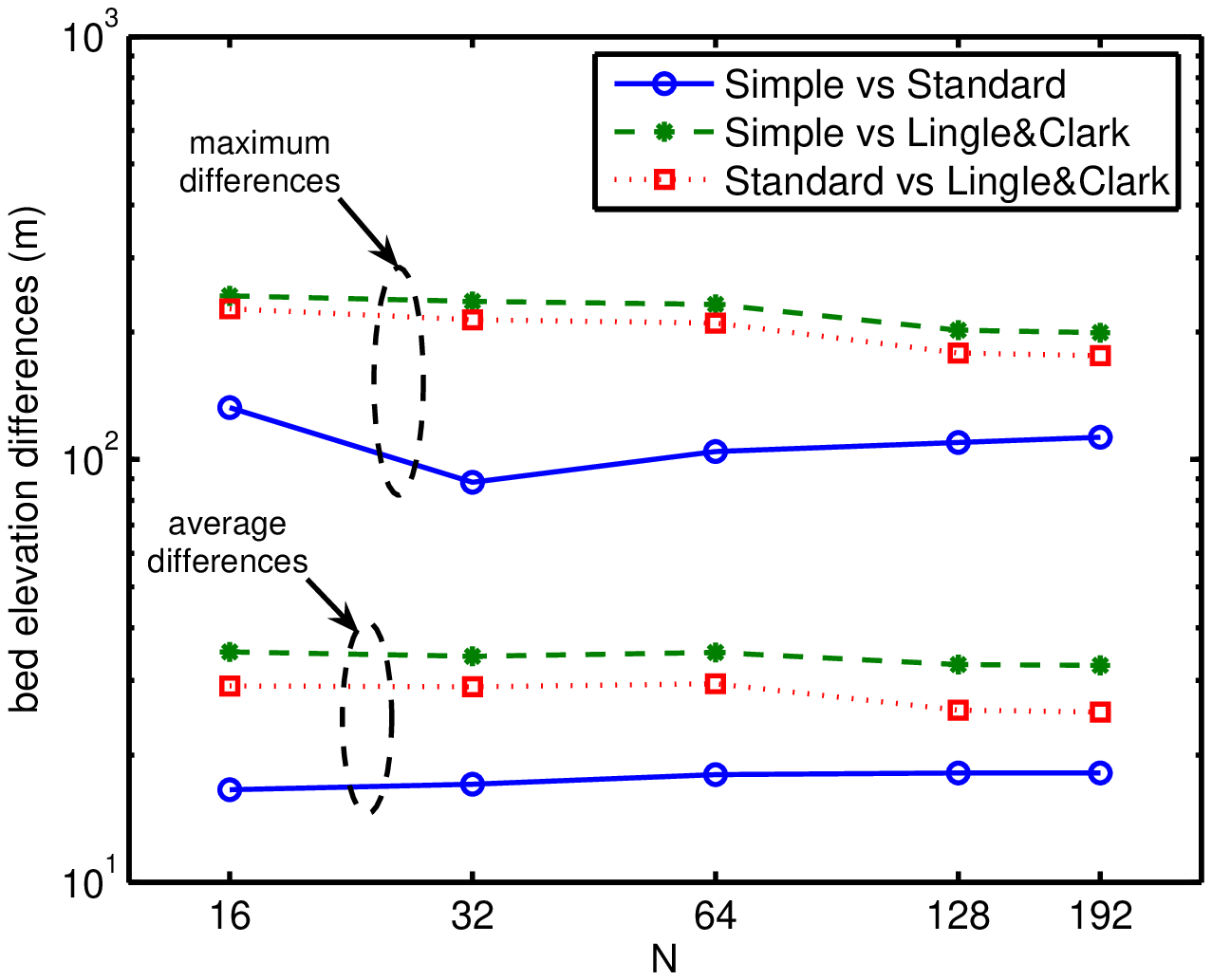}{4.5}
\caption{Maximum and average \emph{bed elevation} differences in pairwise comparison.} \label{fig:beddiffs}
\end{figure}

Now, are these differences significant?  The answer shown in figure \ref{fig:thickdiffsplus} is \emph{yes}.  With a caveat.  As noted in \Bueleretal, ice sheet flow simulations on grids inevitably make large thinkness errors near the margin.  These errors decay only slowly under grid refinement, as can be seen in figure \ref{fig:thickdiffsplus}.

\begin{figure}[ht]
\widefigure{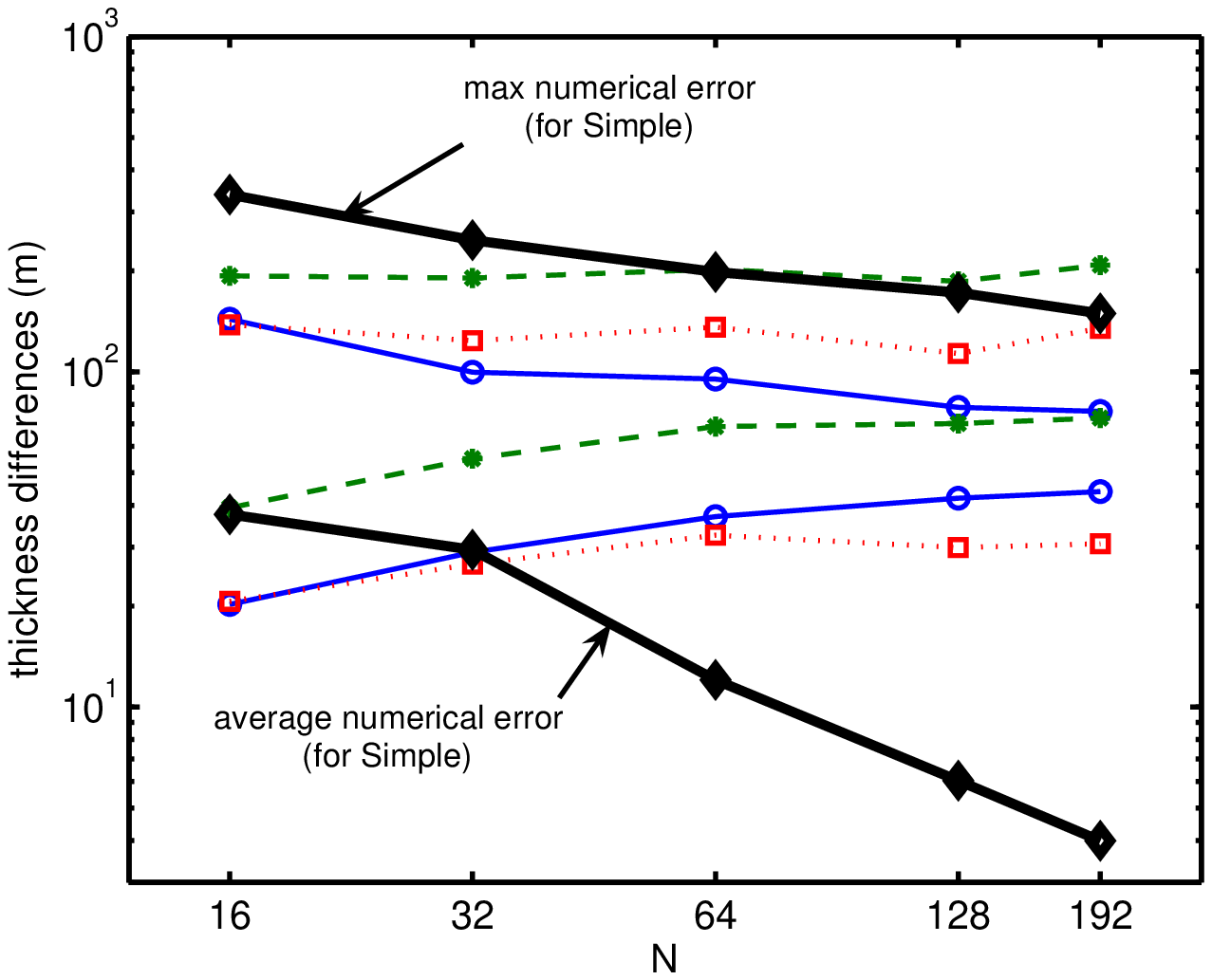}{4.5}
\caption{Ice thickness differences in pairwise comparison as in figure \ref{fig:thickdiffs} but with numerical errors for the simple isostasy case superimposed.  Differences among coupled ice-earth models significantly exceed numerical error except for localized numerical errors within a couple of grid points of the margin.} \label{fig:thickdiffsplus}
\end{figure}

\section{Conclusions} \label{sect:conclude}

We have seen several modeling and computational issues and numerous equations.  So let us identify our major point:  Equation \eqref{pde}
    $$\ppt{}\left(2\eta\,\lap^{1/2}\,u\right) + \rho_r g u +D\lap^2 u = \sigma_{zz},$$
where $\lap$ is the positive Laplacian $\lap=-\partial^2/\partial x^2 - \partial^2/\partial y^2$, is \emph{both}\begin{itemize}
\item a better model for a viscous half space overlain by an elastic plate than the standard model \eqref{notionalplate}, \eqref{notionaldecay} which is widely used in the ice sheet modeling literature,
\item \emph{and} is very computationally tractable on a rectangular grid using a Fourier spectral collocation method.\end{itemize}

In brief, one derives equation \eqref{pde} by starting with equation (4) in \cite{LingleClark}, clearing denominators, and then taking the inverse Hankel transform by recognizing powers of the positive Laplacian $\lap$.  That is, equation \eqref{pde} is equivalent to equation (4) in \cite{LingleClark}.

Of course, equation \eqref{pde} fails to incorporate spherical or self-gravitating effects.  Following \citeasnoun{LingleClark} we have, however, chosen to superpose upon the result of \eqref{pde} a purely-elastic, but spherical and self-gravitating \cite{Farrell}, displacement from equation \eqref{GEuse}.  This is an admittedly \emph{ad hoc} way to incorporate spherical and self-gravitating effects into an earth deformation model.\footnote{The \emph{ad hoc} combination ``$u=u^E+u^V$'' could perhaps be replaced by some other linear combination $u=\alpha u^E + \beta u^V$, $\alpha,\beta>0$, however.}  We note that the effects of equation \eqref{GEuse} are ``longer range'' than those of equation \eqref{pde} at large times, so in some sense it is more important to incorporate ``sphericalness'' into the purely elastic part of the earth model.

\section{Acknowledgements and a Dedication} \label{sect:acknowledge}

Praveena Muthyala and Chris Larsen contributed to its practical and conceptual development, respectively.  It is dedicated to the students in Math 611 in Fall 2005, who suffered from the first author's need to be thorough on the subject of Hankel transforms.

\bibliography{../ice}
\bibliographystyle{agsm}

\appendix
\section{On the Hankel transform and powers of the Laplacian}\label{app:hankel}

A reasonable first reference for the Hankel transform is Chapter 12 of \cite{Bracewell}.  \citeasnoun{Sneddon} is much more complete, however.  Sneddon addresses the application of Hankel transforms to elasticity problems, in particular.

\begin{defn}  Suppose $f(r)$ is defined on $(0,\infty)$ and suppose that $\int_0^\infty |f(r)|\,r\,dr < \infty$.  The \emph{Hankel transform} $\bar f(\kap)$ , $\kap>0$, is
\begin{equation}\label{hankeldefn}
\bar f(\kap) = \int_0^\infty f(r) J_0(\kap r) r\,dr.
\end{equation}
\end{defn}
\noindent This is the transform as normalized by \cite{Sneddon}, \cite{Cathles}, and \cite{LingleClark}; \citeasnoun{Bracewell} is slightly different.

Here $J_0$ is the zeroth-order Bessel function of first kind
    $$J_0(z)=\sum_{k=0}^\infty \frac{(-1)^k}{2^{2k}\,(k!)^2}\, z^{2k},$$
an entire function; note $|J_0(r)|\le 1$ for all $r$.  $J_0(z)$ is the unique solution to the ODE initial value problem
\begin{equation}\label{besselseqn}
z^2 y''(z) + z y'(z) + (z^2-0) y(z)=0, \qquad y(0)=1, \qquad y'(0)=0
\end{equation}
and it has integral formula
\begin{equation}\label{J0int}
J_0(z) = \frac{1}{2\pi} \int_0^{2\pi} e^{-iz\cos\theta}\,d\theta.
\end{equation}
Both \eqref{besselseqn} and \eqref{J0int} will be used below.

The most natural source of the Hankel transform is as the \emph{Fourier transform of a radial function on the plane}.  In particular, suppose $f(x,y)$ is a bounded and integrable function on the plane which is actually radial $f=f(r)$.  Suppose one computes the two-variable Fourier transform $\mathcal{F}_2[f]=\tilde f$ by converting the integral to polar coordinates:
\begin{align*}
\tilde f (\xi,\zeta) &= \frac{1}{2\pi} \int_{-\infty}^\infty \int_{-\infty}^\infty f(r) e^{-i(x\xi+y\zeta)}\,dx\,dy = \int_0^\infty f(r) \left[\frac{1}{2\pi}\int_0^{2\pi} e^{-ir\kap \cos(\theta-\phi)}\,d\theta\right] r\,dr \\
    &= \int_0^\infty f(r) \left[\frac{1}{2\pi}\int_0^{2\pi} e^{-ir \kap\cos\theta}\,d\theta\right] r\,dr = \int_0^\infty f(r) J_0(\kap r) r\,dr
\end{align*}
where $\kap^2=\xi^2+\zeta^2$.  We have used \eqref{J0int} above.  Concretely, we have substituted $x=r\cos\theta$, $y=r\sin\theta$, $\xi=\kap\cos\phi$, and $\zeta=\kap\sin\phi$.  Thus $x\xi+y\zeta=r\kap \cos(\theta-\phi)$, and for fixed $\phi$ the function $\vf(\theta)=e^{-ir\kap \cos(\theta-\phi)}$ is periodic with period $2\pi$.  We conclude that in these circumstances $\tilde f$ is also radial. 

The Hankel transform \eqref{hankeldefn} is evidently linear.  The general two-variable Fourier transform, which we have normalized to be unitary, has the property
    $$\tilde{\tilde g}(x,y)=g(-x,-y).$$
Therefore the map $g \mapsto \tilde{\tilde g}$ is the identity when restricted to the subspace of radial functions, so $\bar{\bar f}(r) = \tilde{\tilde f}(r) = f(r)$ and the Hankel transform is self-inverse.

For $f=f(x,y)$ sufficiently smooth, define the \emph{positive Laplacian} of $f$ to be
    $$\lap f = -\grad^2 f = - \ppxx{f} - \ppyy{f}.$$
We have chosen the sign of the Laplacian to be positive \emph{as an operator}, which is to say that the eigenvalues/spectrum of this operator are nonnegative.  In particular, the Fourier transform of the Laplacian is multiplication by a nonnegative function
    $$\widetilde{\lap f}(\xi,\zeta) = (\xi^2+\zeta^2)\, \tilde f(\xi,\zeta).$$
For a radial function $f(r)$ we recall that $\lap f(r) = -f'' - r^{-1} f'$.

Consider the Hankel transform of the Laplacian.

\begin{lem}
If $f'(r)$ is bounded as $r\to 0^+$ and if $\max\{|f(r)|r, |f'(r)| r\} \to 0$ as $r\to+\infty$ then the Hankel transform formula holds:
    $$\overline{\lap f}(\kap) = \kap^2 \bar f(\kap).$$
\end{lem}

\begin{proof}  Letting prime denote $\partial/\partial r$, two integrations-by-parts do the job:
\begin{align*}
\int_0^\infty \big(-f''(r) &- r^{-1} f'(r)\big) J_0(\kap r) r\,dr \\
    &= f'(r) J_0(\kap r) r\Big]_0^\infty + \int_0^\infty f'(r) \ppr{}\left(J_0(\kap r) r\right)\,dr - \int_0^\infty f'(r) J_0(\kap r)\,dr \\
    &= 0 + \int_0^\infty f'(r) \left[J_0'(\kap r) \kap r\right] \,dr = f(r) J_0'(\kap r) \kap r\,\Big]_0^\infty - \int_0^\infty f(r) \left(J_0'(\kap r) \kap r\right)'\,dr \\
    &\stackrel{\ast}{=} 0 - \int_0^\infty \frac{f(r)}{r} \left(z^2 J_0''(z) + zJ_0'(z)\right)\,dr \stackrel{\dagger}{=} - \int_0^\infty \frac{f(r)}{r} \left(-z^2 J_0(z)\right)\,dr \\
    &=\kap^2 \int_0^\infty f(r) J_0(\kap r) r\,dr = \kap^2 \bar f(\kap).
\end{align*}
We substituted $z=\kap r$ in step $\ast$ to recognize Bessel's equation \eqref{besselseqn} in step $\dagger$.
\end{proof}

It follows that also for powers of the Laplacian we can give nice Hankel transform formulae.  In particular, if
    $$\lap^2 = \grad^4 = \frac{\partial^4}{\partial x^4} + 2\frac{\partial^4}{\partial x^2 \partial y^2} +\frac{\partial^4}{\partial y^4}$$
denotes the biharmonic operator \cite{Sneddon}, and if $f(r)$ is radial and has appropriate boundedness, then
    $$\overline{\lap^2 f}(\kap) = \kap^4 \bar f(\kap).$$

Now we define the \emph{positive square root} $\lap^{1/2}$ \emph{of the Laplacian} $\lap$ via the Fourier transform:\footnote{The Fourier transform gives a spectral resolution of the positive operator $\lap$ acting on the plane, subject to appropriate boundedness at infinity \cite{ReedSimon}.  Thus we are using the functional calculus to define $\lap^{1/2}$.}

\begin{defn}  Define $(\lap^{1/2} f)(x,y)$ by
    $$\widetilde{\lap^{1/2} f}(\xi,\zeta) = \left(\xi^2+\zeta^2\right)^{1/2}\,\tilde f(\xi,\zeta).$$
\end{defn}

Note that $\lap^{1/2}$ acts on scalar functions to produce scalar functions but that it is not a true differential operator.  In particular, it is neither the gradient nor the divergence.  It can, however, be easily calculated using the Hankel transform because the following formula applies for radial functions $f=f(r)$:
    $$\overline{(\lap^{1/2} f)}(\kap) = \kap \bar f(\kap).$$

Finally we consider the Hankel transform of the Dirac delta ``function.''  Care must be taken because the delta function we need is a function only of the radial coordinate while its defining property applies to functions \emph{in the plane}.  In fact, let $\delta_{(x_0,y_0)}(x,y)$ be defined by the integral
    $$\int_{-\infty}^\infty \int_{-\infty}^\infty f(x,y) \delta_{(x_0,y_0)}(x,y) \,dx\,dy = f(x_0,y_0)$$
for all continuous $f(x,y)$.  Let $\delta_0=\delta_{(0,0)}$ denote the delta function at the origin.  In polar coordinates, and for radial functions $f=f(r)$, $\delta_0$ has the property
    $$\int_0^\infty \int_0^{2\pi} f(r) \delta_0(r,\theta) r \,dr\,d\theta = f(0),$$
which simplifies to
\begin{equation}\label{delta0defn}
\int_0^\infty f(r) \delta_0(r) r\,dr = \frac{f(0)}{2\pi}
\end{equation}
because $\delta_0$ is independent of $\theta$.  Thus
    $$\bar{\delta_0}(\kap)=\int_0^\infty \delta_0(r) J_0(\kap r) r\,dr = \frac{J_0(0)}{2\pi}=\frac{1}{2\pi}.$$

\section{The disc load case}\label{app:exactdisc}  As an exercise and for verification purposes we describe a solution to the viscous half-space flat earth equation \eqref{cathlespde}.  The inverse Hankel transform of this solution is a solution to equation \eqref{pde}.

Consider a disc load centered at the map-plane origin, of radius $R_0$ and thickness $H_0$.  In this case
    $$\sigma_{zz}(r) = \begin{cases}-\rho_i g H_0, & 0<r<R_0, \\ 0, & R_0<r\end{cases}$$
where $\rho_i$ is the density of the load;  we might as well suppose it is ice.  Actually, let's suppose this load is applied at time zero and is held in place: $\sigma_{zz}(r,t) = \sigma_{zz}(r) H(t)$.  Because this load is radial, the Hankel-transformed equation \eqref{cathlespde} is useful.  We will assume an undeformed state $\baruV=0$ at time $t=0$.

We need the Hankel transform ${\bar \sigma}_{zz}$:
\begin{align*}
{\bar \sigma}_{zz}(\kappa) &= -\rho_i g H_0 \int_0^{R_0} J_0(\kap r) r\, dr = -\frac{\rho_i g H_0}{\kap^2} \int_0^{\kap R_0} J_0(s) s\, ds = - \rho_i g H_0 R_0 \kap^{-1} J_1(\kap R_0),
\end{align*}
using the change-of-variable $s=\kap r$ and the identity $\frac{d}{ds}\left(s J_1(s)\right) = s J_0(s)$ (formula (8) in Appendix A of \cite{Sneddon}).

Now, we can solve \eqref{cathlespde} because it is simply a collection of decoupled first-order linear ODE problems in time:
    $$\baruV(\kap,t) = \rho_i g H_0 R_0 \frac{\left\{\exp(-\beta(\kap)t/(2\eta\kap))-1\right\}\, J_1(\kap R_0)}{\kap\, \beta(\kap)}$$
for $t>0$ and $\baruV(\kap,t)=0$ for $t\le 0$, where $\beta(\kap)=\rho_r g + D \kap^4$.  The displacement can be found by an (inverse) Hankel transform
\begin{equation}\label{discint}
u^V(r,t) = \rho_i g H_0 R_0 \int_0^\infty \beta(\kap)^{-1} \left\{\exp(-\beta(\kap)t/(2\eta\kap))-1\right\}\, J_1(\kap R_0)\, J_0(\kap r)\,d\kap.
\end{equation}

This integral can be computed numerically, but care must be taken because the integrand is quite oscillatory.  We break up the integrand into many ($\ge 100$) subintervals and call \Matlab's  \mtt{quadl} on each subinterval.  The code \mtt{viscdisc} in Appendix \ref{app:matlab} implements equation \eqref{discint}.  In particular, we believe that computing \eqref{discint} for geophysically reasonable parameter values is quite accurate, and that the result can be used to verify the result from the Fourier spectral collocation method described in section \ref{sect:PDEmethod}.  See section \ref{sect:results} for results.  The displacement for a particular disc load is graphed in figure \ref{fig:discload}.  This solution looks rather like the Green's function $G^V$ plotted in figure \ref{fig:ghv}, as expected, but with a wider depressed area.

The equilibrium limit of this disc load solution is of interest:
\begin{equation}\label{equildisc}
u^\infty(r)=\lim_{t\to\infty} u^V(r,t) = -\rho_i g H_0 R_0 \int_0^\infty \beta(\kap)^{-1}\, J_1(\kap R_0)\, J_0(\kap r)\,d\kap.
\end{equation}
This function satisfies the PDE $\rho_r g u^\infty + D \lap^2  u^\infty = \sigma_{zz}$.  It appears in figure \ref{fig:discload}, and we see the peripheral ``dip'' and ``bulge'' which occur at the edge of the disc.  Note that the central part of disc load of thickness $H_0$ and large radius descends to the ``compensation depth'' $-(\rho_i/\rho_r)H_0$.

\begin{figure}[ht]
\widefigure{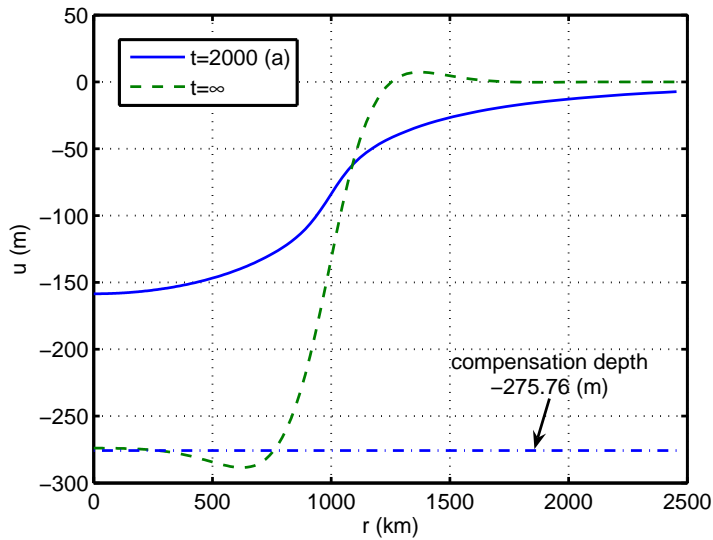}{4.0}
\caption{Vertical displacement at $2000$ years, and the equilibrium position at time $\infty$, for a disc load of ice with thickness $1000$ (m) and radius $1000$ (km).  ``Compensation depth'' corresponding to $\rho_i=910$, $\rho_r=3300$ ($\text{kg}\, \text{m}^{-3}$) also shown.  Note log scale on horizontal axis.} \label{fig:discload}
\end{figure}

\section{On ``Green's function thinking'' for PDE \eqref{pde}} \label{app:greenthinking}  Here we make some comments on formula \eqref{GHVuse} in section \ref{sect:twolin} which relate it to results in section \ref{sect:PDEmethod}.  The time integral in \eqref{GHVuse} starts from $-\infty$ to avoid requiring precise knowledge of the displacement $u^V(x,y,t_0)$ at any particular past time $t_0$.  That is, \eqref{GHVuse} computes the displacement at time $t$ caused by a load history known so far into the past that prior displacement states are irrelevant.  This is reasonable because the underlying PDE, equation \eqref{pde} in section \ref{sect:PDEmethod}, is diffusive and thus the influence of any displacement state decays exponentially in time.

If, on the other hand, an initial displacement $u(x,y,t_0)$ is known at a relatively recent past time then we must adjust \eqref{GHVuse}.  As a start, integrating \eqref{GHVuse} by-parts in the time variable and assuming $\Psi(x,y,-\infty)$ is zero gives the formula
\begin{align}
u^V(x,y,t) &= \int_{-\infty}^t \iint\limits_R \ppt{G^V}(|\br-\br'|,t-t') \Psi(x',y',t') \,dx'\,dy'\,dt' \label{GHVibp}.
\end{align}
Our assumption that there is no load at $t=-\infty$ is equivalent to assuming $\Psi(x,y,t)= \int_{-\infty}^t \lam(x,y,x)\,ds$.  The significance of \eqref{GHVibp} is that the load function $\Psi$ reappears and that we are motivated to examine $\partial G^v/\partial t$.

From \eqref{ghv} note that
\begin{equation}\label{dGHVdt}
\ppt{G^V}(r,t) = - \frac{g}{2\pi} \int_0^\infty (2\eta\kap)^{-1} \exp\left[-\beta(\kap) t/(2\eta\kap)\right]\,J_0(\kap r)\,\kap\,d\kap.
\end{equation}
In fact, formula \eqref{dGHVdt} can also be extracted from equation \eqref{fouriersoln} in section \ref{sect:PDEmethod}:
\begin{align}
\ppt{G^V}(x,y,t) &= -\frac{g}{2\pi} \mathcal{F}_2^{-1} \left\{ \frac{\exp\left[-\beta(\xi,\zeta)t /(2\eta(\xi^2+\zeta^2)^{1/2})\right]}{2\eta (\xi^2+\zeta^2)^{1/2}} \right\} \label{dGVdtfour} \\
    &= -\frac{g}{2\pi} \int_0^\infty (2\eta \kap)^{-1}\,\exp\left[-\beta(\kap)t /(2\eta\kap)\right]\, J_0(\sqrt{x^2+y^2}\,\kap)\,\kap\,d\kap. \notag
\notag\end{align}
Here $\mathcal{F}_2$ stands for the two-variable Fourier transform and $\beta(\kap)=\rho_r g + D \kap^4$ as in section \ref{sect:twolin}.  The second equality in \eqref{dGVdtfour} is explained by noting (Appendix \ref{app:hankel}) that $\mathcal{F}_2$ becomes the Hankel transform on radial functions.  Comparing to section \ref{sect:twolin}, $\sigma_{zz}(x,y,t) = -g \Psi(x,y,t)$ ($\text{N m}^{-2}$) if $\Psi$ ($\text{kg m}^{-3}$) is the load function.

Returning now to equation \eqref{GHVibp}, we can see from \eqref{fouriersoln} how to modify \eqref{GHVibp} to include knowledge of displacement at a finite time $t_0$.  In fact, the convolution theorem
    $$\mathcal{F}_2^{-1}\left\{\tilde f \,\tilde g\right\} = \frac{1}{2\pi} (f\ast g) = \frac{1}{2\pi} \iint\limits_{\phantom{A}\RR^2} f(x-x',y-y')g(x',y')\,dx'\,dy'$$
now allows us to write the inverse Fourier transform of \eqref{fouriersoln} as
\begin{align}
u^V(x,y,t) &= \int_{t_0}^t \iint\limits_{\phantom{A}\RR^2} \ppt{G^V}(|\br-\br'|,t-s) \Psi(x',y',s)\,dx'\,dy'\,ds \label{GVagain} \\
    &\qquad\qquad + \iint\limits_{\phantom{A}\RR^2} \gamma(|\br-\br'|,t-t_0) u(x',y',t_0)\,dx'\,dy' \notag
\end{align}
where
    $$\gamma(r,t) = \frac{1}{2\pi} \int_0^\infty \exp\left[-\beta(\kap)t /(2\eta\kap)\right]\, J_0(r\,\kap)\,\kap\,d\kap.$$

In our view equation \eqref{GVagain} proves equation \eqref{GHVuse}.  We have justified the heuristic Green's function thinking in section \ref{sect:twolin} by a standard linear analysis of PDE \eqref{pde}.  Note that as $t\to\infty$, $\gamma(r,t)\to 0$.  Thus as $t_0\to-\infty$ the last term in \eqref{GVagain} disappears.  The decaying kernel $\gamma$ describes the rate at which information held in the previous displacement $u(x,y,t_0)$ is eliminated.

\section{\Matlab codes} \label{app:matlab}  This section gives \Matlab implementations of the numerical strategies outlined in sections \ref{sect:computelr}, \ref{sect:PDEmethod}, and Appendix \ref{app:exactdisc}.  Here is a synopsis of the three codes:

\begin{itemize}
\item \mtt{geforconv} precomputes the spherical elastic load response matrix $I^E$, given by equation \eqref{Ipq}, which is the computational form of the Green's function $G^E$.  It numerically computes integrals \eqref{Ipq} by using \Matlab's \mtt{dblquad} with the default settings, where the integrand comes from linearly interpolating the tabular data given by \cite{Farrell}.  If \mtt{geforconv} has been run on the given grid, \mtt{fastearth} (below) then convolves $I^E$ with the load, as noted, using \Matlab's \mtt{conv2} for now because it seems sufficiently fast; an FFT-based alternative is reasonable.  The precomputation is much more expensive than running the viscous deformation model or convolving $I^E$ with the load.  For instance, in the \Matlab run
\beginV
tic, I = geforconv(256,256,2000,2000); toc
save I256 I
tic, fastearth(256,100000,500,4,'testC'); toc
\end{verbatim}
\Vend
the call to \mtt{geforconv} requires about 4 \emph{hours} while only 10 \emph{minutes} is required to run \mtt{fastearth}; the machine is a 3 GHz Pentium IV.
\item \mtt{fastearth} computes the viscous model using \mtt{fft2} to compute equation \eqref{DFTpde}.  It also convolves $I^E$ with the load for the spherical elastic model as noted.  In this example implementation the load is either a standard disc load or a modification to  Test C from \Bueleretal.
\item \mtt{viscdisc} implements formula \eqref{discint} in Appendix \ref{app:exactdisc}.
\end{itemize}

\bigskip\noindent\large{\mtt{geforconv.m}:}
\Vfile{geforconv.m}

\bigskip\noindent\large{\mtt{fastearth.m}:}
\Vfile{fastearth.m}

\bigskip\noindent\large{\mtt{viscdisc.m}:}
\Vfile{viscdisc.m}

\end{document}